\pdfoutput=1   % for arXiv submission, ensures PDF output, set a word size limit
\documentclass[journal, 8pt]{IEEEtran} 
\usepackage[utf8]{inputenc} % allow utf-8 input
\usepackage[T1]{fontenc}    % use 8-bit T1 fonts
\usepackage{hyperref}       % hyperlinks
\usepackage{url}            % simple URL typesetting
\usepackage{booktabs}       % professional-quality tables
\usepackage{amsfonts}       % blackboard math symbols
\usepackage{nicefrac}       % compact symbols for 1/2, etc.
\usepackage{microtype}      % microtypography
\usepackage{lipsum}
\usepackage{fancyhdr}       % header
\usepackage{graphicx}       % graphics
\usepackage{geometry}       % to set the page margins
\usepackage{caption}
\usepackage{graphicx} 
\usepackage{subfigure}
\usepackage{amsmath}
\usepackage{amssymb}
\usepackage{overpic}
\usepackage{enumitem}
\usepackage{amsthm}
\usepackage[noend]{algpseudocode}
\usepackage{algorithmicx,algorithm}
\usepackage{algorithm}
\usepackage{scalerel,stackengine}
\usepackage{multirow}
\usepackage{xcolor}
\usepackage{tikz}
\usepackage{tikz-3dplot}
\usepackage[numbers]{natbib}% enables natbib package, with numerical citation style
\usepackage[pagewise, switch]{lineno}  % 启用分页重新编号

\begin{document}
% \linenumbers  % 启用行号

\title{Meta-DSP: A Meta-Learning Approach for Data-Driven Nonlinear Compensation in High-Speed Optical Fiber Systems}

\author{Xinyu Xiao\thanks{Xinyu Xiao, School of Mathematical Science, Peking University. xiao-xin-yu@pku.edu.cn}, 
Zhennan Zhou\thanks{Zhennan Zhou, Corresponding author. Institute for Theoretical Sciences, Westlake University, Hangzhou, 310030, China (zhouzhennan@westlake.edu.cn).}, 
Bin Dong\thanks{Bin Dong, Corresponding author. Beijing International Center for Mathematical Research, Peking University; Center for Machine Learning Research, Peking University; National Biomedical Imaging Center, Peking University. dongbin@math.pku.edu.cn},
Dingjiong Ma\thanks{Dingjiong Ma, Theory Lab, Central Research Institute, 2012 Labs, Huawei Technology Co. Ltd., Hong Kong, China. ma.dingjiong2@huawei.com}, 
Li Zhou\thanks{Li Zhou, Theory Lab, Central Research Institute, 2012 Labs, Huawei Technology Co. Ltd., Shanghai,P.R. China. zhouli107@huawei.com}, 
Jie Sun\thanks{Jie Sun, Theory Lab, Central Research Institute, 2012 Labs, Huawei Technology Co. Ltd., Hong Kong, China. j.sun@huawei.com}.}

% \markboth{Journal of LightWave Technology,~Vol.~14, No.~8, June~2025}% 

\maketitle

\begin{abstract}
Nonlinear effects in high-speed optical fiber systems fundamentally limit channel capacity. While traditional Digital Backward Propagation (DBP) with adaptive filters addresses these effects, its computational complexity remains impractical. Data-driven solutions like Filtered DBP (FDBP) reduce complexity but critically lack inherent generalization: Their nonlinear compensation capability cannot be naturally extended to new transmission rates or WDM channel counts without retraining on newly collected data.
We propose Meta-DSP, a novel signal processing pipeline combining: (1) Meta-DBP, a meta-learning-based DBP model that generalizes across transmission parameters without retraining, and (2) XPM-ADF, a carefully engineered adaptive filter designed to address multi-channel nonlinear distortions. The system demonstrates strong generalization, learning from 40 Gbaud single-channel data and successfully applying this knowledge to higher rates (80/160 Gbaud) and multi-channel configurations (up to 21 channels). Experimental results show Meta-DSP improves Q-factor by 0.55 dB over CDC in challenging scenarios while reducing computational complexity 10$\times$ versus DBP. This work provides a scalable solution for nonlinear compensation in dynamic optical networks, balancing performance with practical computational constraints.
\end{abstract}


\section{Introduction}

Optical communication networks carry the most traffic in communication networks. In recent years, with the surge in network traffic, optical communication networks have been transitioning towards higher speeds, greater distances, and larger capacities. Wavelength division multiplexing (WDM) technology is a crucial technique for expanding optical fiber capacity. WDM modulates the information of multiple channels onto different optical frequency carriers and integrates them onto the same fiber for propagation, greatly enhancing the propagation capacity of commercial systems. However, WDM systems face a fundamental trade-off: while increasing signal power improves the signal-to-noise ratio (SNR), it simultaneously induces severe nonlinear noise through optical fiber Kerr effects. These nonlinear impairments take two primary forms: self-phase modulation (SPM), caused by intra-channel nonlinear effects, and cross-phase modulation (XPM), resulting from inter-channel nonlinear interactions.

Digital Signal Processing (DSP) techniques have been developed to mitigate various interferences and noise in optical communication. The DSP module in the receiver addresses transmission distortions such as chromatic dispersion, polarization mode dispersion, carrier phase noise, and nonlinear Kerr effects. This paper investigates an advanced DSP architecture for nonlinear compensation, featuring a cascaded structure that combines DBP for dispersion and SPM compensation with an Adaptive Filter (ADF) for XPM mitigation.

DBP is widely recognized as one of the most effective algorithms for compensating SPM \cite{liga2014multichannelDBP}. This technique compensates for both linear and nonlinear interference by solving the inverse transmission equation of light waves. However, the highly heterogeneous nature of DBP makes its implementation challenging. The number of backward steps required in high-speed, long-distance systems can be daunting \cite{2010Rs-DBP}.  While perturbation-based compensation (PBC) \cite{tao2011multiplier} offers a computationally simpler alternative, its nonlinearity mitigation efficiency falls significantly short of DBP. With the emergence of deep learning technologies in recent years, an increasing number of studies have leveraged data-driven methods in optical communications, where model parameters are learned directly from data to significantly reduce computational complexity. A significant amount of works have related the structure of DBP to deep neural networks\cite{fan2020NC},\cite{fan2021combined}, \cite{hager2018nonlinear},\cite{CNNDBP}, considerably reducing the number of steps required for DBP. However, existing studies about learnable DBP have only considered a single system parameter setting: fixed transmission rates and WDM channel numbers. Their scalability under these system parameters is poor, requiring model retraining for different settings. For the same optical channel, its nonlinear physical properties should remain constant. Therefore, an effective learnable DBP scheme should be able to adapt to different system parameters without requiring multiple retrainings. This is one of the key motivations behind our approach.

ADF is a dynamic signal processing technique widely employed in optical communication systems, with typical applications including Frequency Offset Estimation (FOE), Carrier Phase Estimation (CPE), and Multiple-Input Multiple-Output (MIMO) processing. By dynamically adjusting filter parameters, ADF effectively mitigates noise and enhances system performance. Recent studies have shown that XPM, due to its inherent temporal correlation, can also be partially captured by adaptive filters \cite{2016VL},\cite{dar2013properties},\cite{secondini2012analytical}. Significant progress has been made in this research direction. For instance, Recursive Least Square (RLS) algorithm has been employed for XPM compensation in \cite{dar2014inter,nassaji2023xpm,lian2020xpm}. The work in \cite{CNNDBP} proposed using Least Mean Square (LMS) algorithm to compensate for the nonlinear phase noise that DBP fails to address. Considering the phase-dominant nature of XPM noise \cite{dar2017nonlinear}, classical phase estimation algorithms such as Blind Phase Search (BPS) and Viterbi-Viterbi (V\&V) algorithm have also been investigated \cite{zheng2019xpm}. Nevertheless, these methods primarily repurpose existing algorithms without structural modifications tailored to XPM's unique spatiotemporal properties, highlighting the need for specialized XPM compensation architectures.

\begin{figure}[ht]
    \centering
    \includegraphics[width=0.45\textwidth]{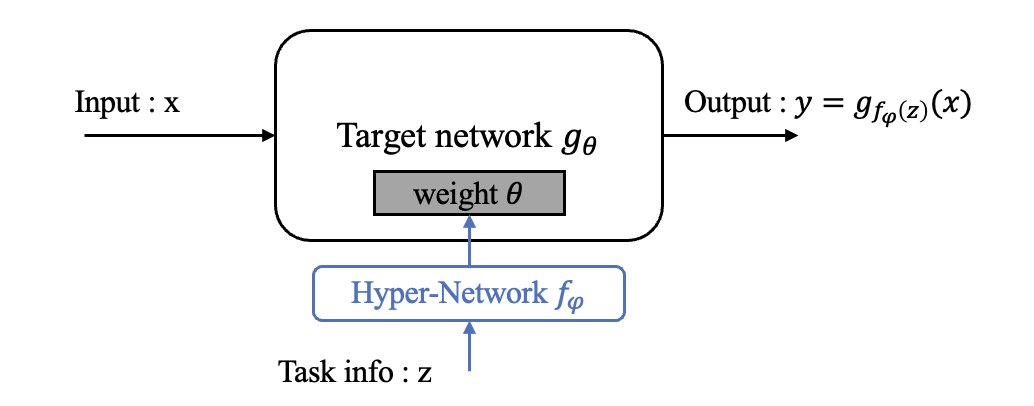}
    \caption{Hypernetwork is a neural network that generates the weights of another network, typically referred to as the target network. By learning the mapping from task encoding to network weights, hypernetworks enable dynamic weight generation, model adaptivity, and parameter efficiency, which are particularly beneficial in scenarios with diverse tasks or resource constraints.}
    \label{hyper_illustration}
\end{figure}

In this paper, we propose Meta-DSP, a meta-learning-inspired DSP model composed of Meta-DBP and XPM-ADF, which is capable of concurrently handling multimodal data from optical fibers.  Here, multimodal refers to the variety of optical signal data collected from the same optical fiber link, ranging from low-speed to high-speed, low-power to high-power, and with varying numbers of WDM channels. Meta-DBP is a learnable DBP module capable of automatically transferring nonlinear compensation knowledge across different system parameters to mitigate nonlinear interference, specifically SPM, within the optical channel. To achieve this, we employ a technique called a hypernetwork \cite{2016hypernetworks}. A hypernetwork $f_{\phi}(z)$ is a neural network that generates the weights for another network, typically referred to as the target network $g_{\theta}(x)$ as shown in Figure \ref{hyper_illustration}. By learning the mapping from task conditions to network weights $\theta = f_{\varphi}(z)$, hypernetworks enable dynamic weight generation, model adaptability, and parameter efficiency, which are especially useful in scenarios involving diverse tasks or resource constraints. 
XPM-ADF is an adaptive filter designed to efficiently capture nonlinear interference between channels. It integrates a finite impulse response (FIR) filter and a single-tap phase estimator, using the Normalized Least Mean Square (NLMS) algorithm for updates. Simulation results show that this module effectively captures XPM interference.

We train Meta-DSP with supervised learning on simulated data 40 Gbaud and single channel. The properly trained Meta-DSP shows an increase of 0.55 dB in the signal's Q factor compared to CDC on the most challenging scenario of 160 Gbaud, 21 channels. Meta-DSP also offers a tenfold reduction in computational complexity compared to DBP while achieving a similar Q factor level. The architecture of Meta-DSP holds universal significance in the algorithmic design of communication systems. It can be integrated into various stages of a comprehensive communication system, bolstering its adaptability and generalization capabilities. Our research underscores the potential of the meta-learning-based digital signal processing module across pivotal parameters in optical communication frameworks.

The rest of the paper is organized as follows: 
Section 2 introduces Meta-DSP and analyzes the training and testing details of the model. 
Section 3 introduce our simulated transmission system model.
Section 4 reports the results of our numerical simulation results. Finally, 
Section 5 discusses future directions for our compensation algorithm.

\section{Proposed method: Meta-dsp}
\begin{figure*}[htbp]
    \centering
    \includegraphics[width=0.98\textwidth]{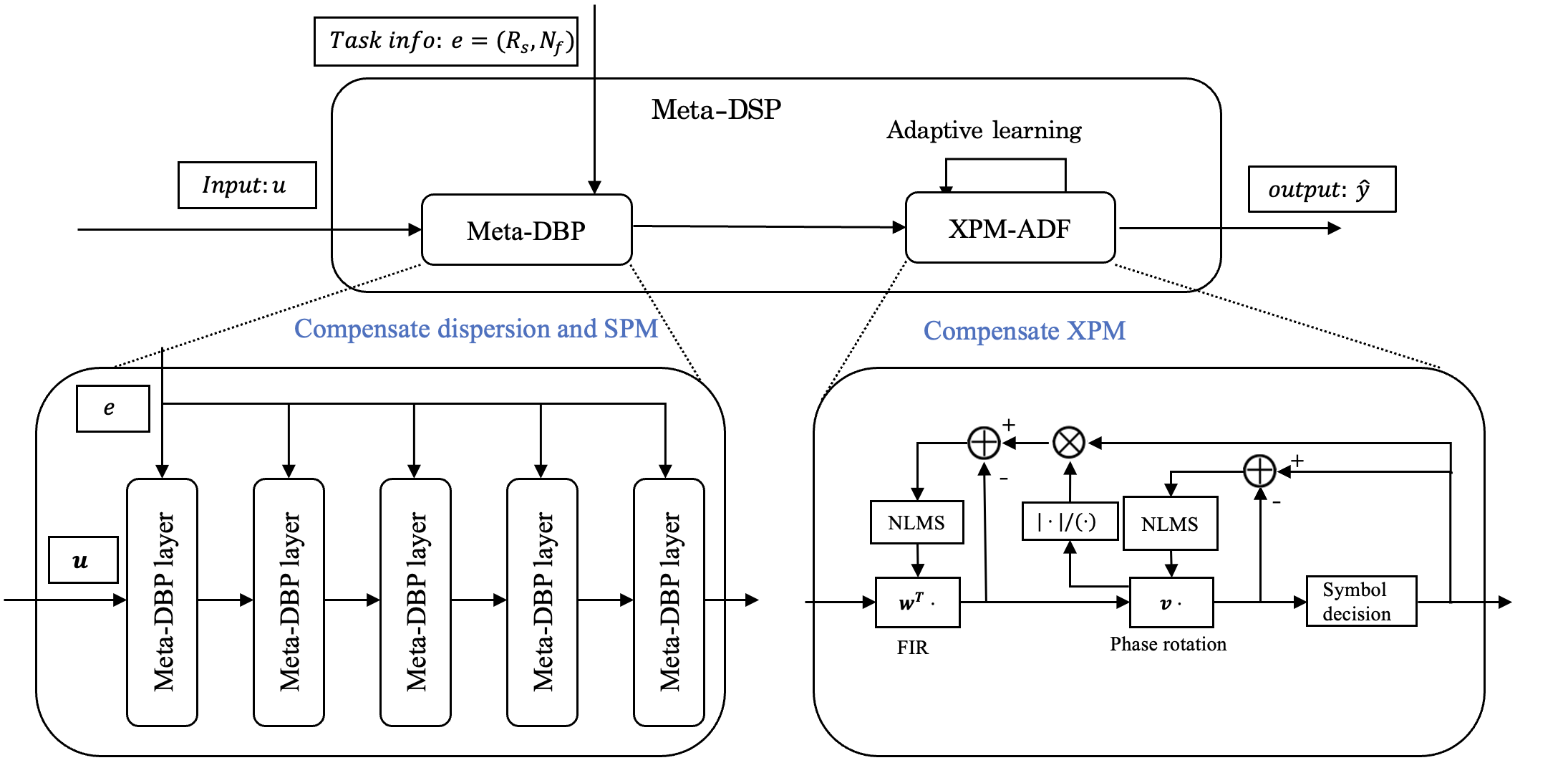}
    \caption{Schematic of the Meta-DSP. The system comprises the Meta-DBP and XPM-ADF. Meta-DBP compensates for dispersion and SPM at a sample rate of 2 samples per symbol under the task-specific information. XPM-ADF capture the temporal statistical characteristics of XPM noise using a FIR filter $\mathbf{w}$ and a phase estimator $\mathbf{v}$. $\mathbf{u}$: input signal. $e$: task information include symbol rate $R_s$, length of N-filter $N_f$. $\hat{y}$: output signal.}
    \label{MetaDSP}
\end{figure*}

In this section, we propose Meta-DSP, a meta-learning based digital signal processing model. 
Classical Learnable DBP (LDBP) have poor adaptivity to multi-modal fiber data due to their inherent static structure. 
In our work, we propose improving classical DSP with insights and techniques from meta-learning. 
The core idea of meta-learning is training a model that can quickly adapt to new tasks \cite{vanschoren2018meta}. In the context of optical communication, this new task involves varying transmission rates, power levels, number of channels. We aim to design a nonlinear compensation model that can learn sufficient nonlinear compensation knowledge from fiber data under a single setting. This would enable the model to perform well across a wide range of conditions, including different transmission rates, power levels, and channel configurations.
Specifically, Meta-DSP consists of two modules: Meta-DBP and XPM-ADF. The whole structure of Meta-DSP is shown in Fig \ref{MetaDSP}.
The function of Meta-DBP is to compensate for the deterministic nonlinear effect SPM within the channel, while the function of XPM-ADF is to compensate for the stochastic nonlinear effect XPM between channels.

The first module, Meta-DBP, is a meta-learning version of LDBP. Past works on LDBP \cite{hager2018nonlinear, hager2020, fan2020NC, fan2021combined, CNNDBP, PA-DBP} have all adopted the approach of inserting a learnable low-pass filter into the nonlinear operator to replace the traditional nonlinear phase rotation. This method is also referred to as Filtered DBP (FDBP) in the literature. These models learn to compensate for the nonlinear effects in the channel by learning appropriate filter coefficients.
However, the filter coefficients in these models are fixed, and once the transmission scenario changes, such as altering the transmission rate, the performance of these models drops sharply, necessitating retraining. To obtain an efficient DBP model that adapts to different parameter settings, we enhance the FDBP model using hypernetwork technology. We use the FDBP model as the target network and employ a hypernetwork to generate the weights in the FDBP based on task-specific information, thereby enhancing the model's generalization capability.

The second module, XPM-ADF, is a specialized design of ADF to compensate for XPM noise. XPM noise arises from the interference of optical power from other channels on the target channel, making it a random nonlinear noise. Fortunately, previous studies \cite{2016VL}, \cite{dar2013properties}, \cite{secondini2012analytical} have identified temporal correlations in XPM noise. As a result, although XPM cannot be fully compensated due to information loss, its statistical properties can be effectively captured by a well-designed adaptive filter, which can help mitigate the interference caused by XPM noise. In our study, we use a FIR filter and a phase estimator to capture XPM noise. The inclusion of the phase estimator allows our model to track the phase variations of XPM noise more quickly, resulting in more accurate and stable error signals for updating the FIR filter.

In this section, we first introduce the FDBP algorithm, which is the foundation of our Meta-DBP. 
Then, we introduce the Meta-DBP and XPM-ADF in detail. 
Finally, we present the training and testing process of Meta-DSP. 

\begin{figure*}[htbp]
    \centering
    \includegraphics[width=0.8\textwidth]{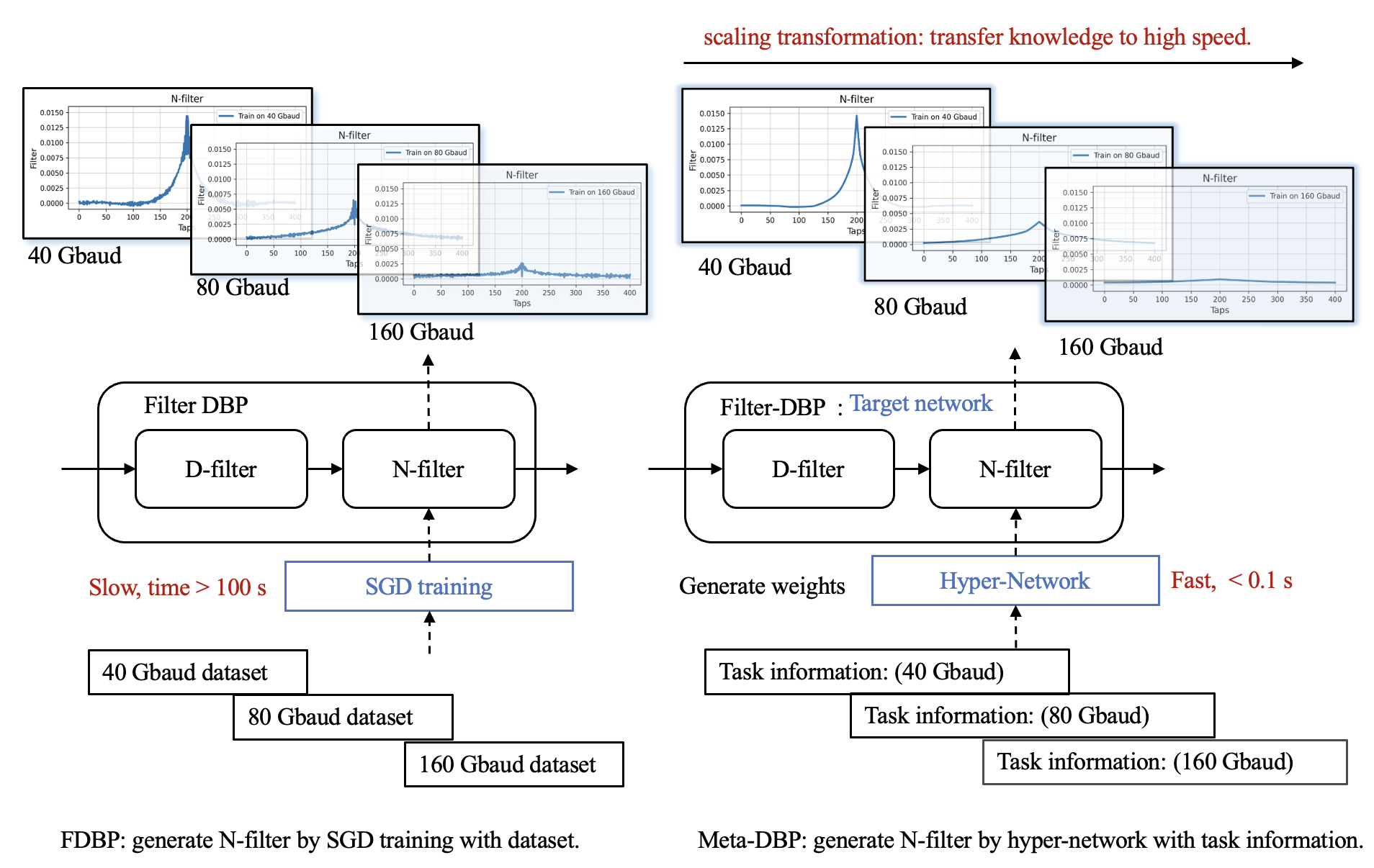}
    \caption{Meta-DBP use a hyper-network to inference N-filter while FDBP get N-filter using SGD traning each time. FDBP requires collecting a new dataset for different symbol rates and training with SGD for an extended period (several minutes) to obtain the parameters. In contrast, Meta-DBP uses a hyper-network to learn the complete nonlinear compensation knowledge from 40 Gbaud data and transfers this knowledge to 80 Gbaud and 160 Gbaud. Once the hypernetwork is properly trained, the N-filter can be inferred at any symbol rate with ultra-fast speed(less than 0.1 seconds).}
    \label{FDBP vs Meta-DBP}
\end{figure*}

\subsection{Filterd DBP}
DBP is a algorithm based on the Nonlinear Schrödinger equation (NLSE) for single polarizarion system. 
\begin{equation}\label{NLSE}
    \begin{split}
    &\frac{\partial u(z, t)}{\partial z}=-\frac{\alpha}{2}u(z,t) + \frac{i\beta_2}{2} \frac{\partial^2 u(z,t)}{\partial t^2} \\
    & - i \gamma \left|u(z, t)\right|^2 u(z, t) + n(z,t).
    \end{split}
\end{equation}
While the dual-polarization form, commonly referred to as the Manakov equation, is given as follows:
\begin{equation}
\begin{split}
&\frac{\partial u_{x/y}(z, t)}{\partial z}=-\frac{\alpha}{2}u_{x/y}(z,t) + \frac{i\beta_2}{2} \frac{\partial^2 u_{x/y}(z,t)}{\partial t^2} \\
& -  \frac{i 8\gamma}{9} \left(\left|u_{x}(z, t)\right|^2 + \left|u_{y}(z, t)\right|^2 \right)u_{x/y}(z, t).
\end{split}
\end{equation}
Here, $u(z,t)$ represents the optical pulse in a fiber at position $z$ and time $t$, and $n(z, t)$ denotes the distributed amplified spontaneous emission (ASE) noise from inline amplifiers. The noise in \eqref{NLSE} can be modeled as additive white Gaussian noise (AWGN) with zero mean and zero auto-correlation. The parameters $\alpha$, $\beta_2$, and $\gamma$ represent attenuation, group velocity dispersion, and the fiber nonlinearity coefficient, respectively. The electric field of a signal at the transmitter is represented by $u(0,t)$, while $u(L,t)$ denotes the received signal at the receiver, neglecting the carrier frequency offset and carrier noise.

To compensate for distortions induced by fiber propagation, the DBP method aims to reverse the propagation process by performing the inverse of the split-step Fourier method. It iteratively processes the signal through multiple backpropagation steps, each consisting of a dispersion compensation module and a nonlinear phase correction module.
Dispersion compensation module can be implemented in the time domain use convolution operation $\mathbf{u} * \mbox{D-filter}$, the dispersion kernel "D-filter" is given as:
\begin{align}\label{D-eq-time-dis}
    & \mbox{D-filter} = \mbox{ifft}\left(\exp\left(-\frac{i\beta_2 \omega^2}{2} \cdot dz\right)\right)  \nonumber \\
    & \omega = 2\pi F_s \cdot \frac{1}{N_d}  \left[0,1,\ldots,-\frac{N_d+1}{2},-\frac{N_d-1}{2},\ldots,-1\right],
\end{align}
Where $F_s = R_s \times \mbox{SpS}$ is the sample rate of signal, $\mbox{SpS}$ is samples per symbol, and $dz$ is the step size of the backpropagation, $N_d$ denotes the length of the D-kernel and 
should not be less than the sample distance affected by dispersion. An empirical expression for
this has been provided in numerous literature \cite{CNNDBP}.
\begin{equation}
    N_d = [2 \pi dz \beta_2 F_s^2].
\end{equation}
Nonlinear compensation module is implemented as a phase rotation operation:
\begin{equation}\label{N-eq-dis}
     \exp(-i\gamma dz |\mathbf{u}|^2) \mathbf{u},
\end{equation}
To enhance DBP's performance, FDBP has been proposed, which assumes a low-pass filter for the signal power to achieve an improved phase rotation at the nonlinear step \cite{2010improvedDBP},  \cite{ESSFM}, \cite{CNNDBP}, \cite{fan2021combined}. 
FDBP expands the nonlinear phase rotation in \eqref{N-eq-dis} as follows:
\begin{equation}\label{FDBP}
     \exp(i\gamma dz |\mathbf{u}|^2 * \mbox{N-filter}) \mathbf{u},
\end{equation}
Where $\mbox{N-filter}$ refers to the model parameters that can be optimized during the model's training phase, and $*$ in \eqref{FDBP} denotes a central convolution. In the case of single polarization, the N-filter is a convolution kernel with shape \([N_f]\). For dual polarization, the N-filter becomes a convolution kernel of shape \([2, 2, N_f]\), represented as \( \mbox{N-filter} = [[N_{xx}, N_{xy}], [N_{yx}, N_{yy}]] \). Here, the \( * \) denotes a convolution operation where both the input and output channels are 2. It has been validated in the literature that FDBP can reduce the computational cost of DBP by an order of magnitude.

\subsection{Meta-DBP}
Meta-DBP uses a hypernetwork $f_{\varphi}$ to generate the N-filter for nonlinear phase rotation in FDBP, as shown in Figure \ref{FDBP vs Meta-DBP}. Unlike FDBP, which obtains the N-filter through stochastic gradient descent on different modal datasets, Meta-DBP directly infers the N-filter using a hypernetwork $f_{\varphi}$ based on task-specific information (symbol rate $R_s$), making the process more direct and efficient. To enable the hypernetwork \(f_{\varphi}\) to effectively transfer nonlinear compensation knowledge across different modalities, we investigate the relationships between N-filters in different modalities through both theoretical derivation and simulation experiments. We find that N-filters exhibit parallel transferability across different WDM channel counts and power levels. Additionally, N-filters under varying transmission baud rates are approximately related by a scaling transformation. Finally, we incorporate these findings into the design of our hypernetwork.
\begin{figure*}[htbp]
    \centering
    \includegraphics[width=0.8\textwidth]{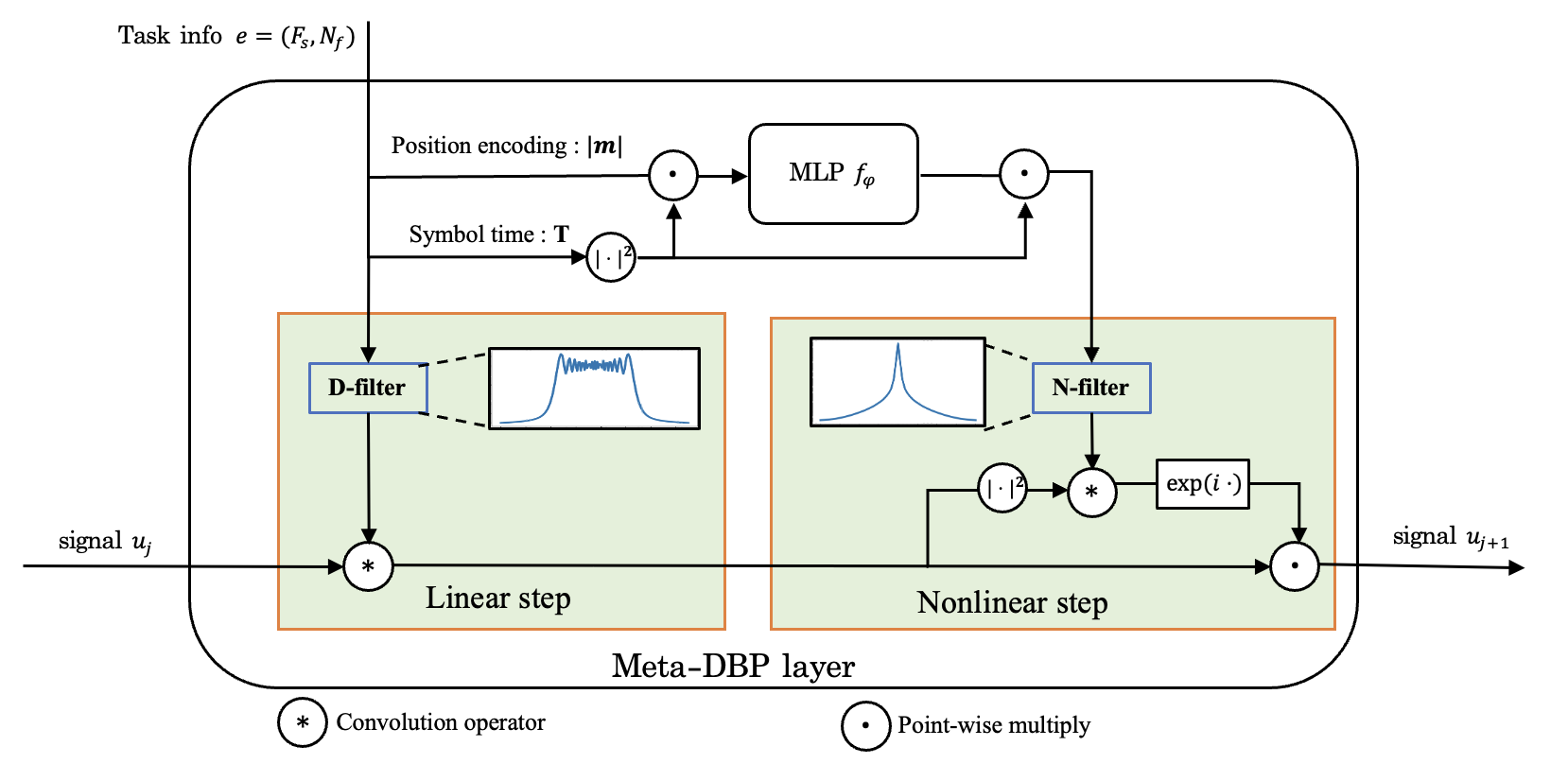}
    \caption{Meta-DBP layer. Meta-DBP use a Multi-layer Perceptrons (MLP) to inference N-filter for nonlinear phase rotation, while FDBP regard N-filter as a static parameter.
    $*$ is convolution operation, $\cdot$ is point-wise multiplication.}
    \label{Meta-DBP}
\end{figure*}

We first analyze the relationship between the N-filter and task-specific information from a theoretical perspective. In simple terms, the N-filter is essentially equivalent to the nonlinear coefficient $C_{m,0}$ of the intral-channel XPM term in the Perturbation Based Compensation (PBC)\cite{lin2021perturbation}. To better illustrate this relationship, we first review the PBC:
\begin{align}
    a_{p,x} + i\sum_{m,n} C_{m,n} \left( a_{p+m,x} a_{p+m+n,x}^* + a_{p+m,y} a_{p+m+n,y}^* \right) a_{p+n,x}, \nonumber \\
    a_{p,y} + i\sum_{m,n} C_{m,n} \left( a_{p+m,x} a_{p+m+n,x}^* + a_{p+m,y} a_{p+m+n,y}^* \right) a_{p+n,y},
\end{align}
where $a_{p,x}$ and $a_{p,y}$ represent the received symbols in the two polarization directions, $C_{m,n}$ is the nonlinear coefficient which can be approximated by the following formula:
\begin{align}
    C_{m,n} &= \frac{1}{T} \int_0^z dz \,  \gamma f(z) \int dt \, \Big[ g^*(z,t) g(z,t-mT) \nonumber \\
            & \quad \times g(z,t-nT) g^*(z,t-(m+n)T) \Big],
\end{align}  
where $g(z,t)$ is the pulse function after dispersion evolution over distance $z$. If we retain only the terms about intral-channel XPM term on the right-hand side, the PBC method can be simplified as:
\begin{align}
    a_{p,x} + i\sum_{m \neq 0} C_{m,0} \left( 2|a_{p+m,x}|^2 + |a_{p+m,y}|^2 \right) a_{p,x} \nonumber \\ 
    a_{p,y} + i\sum_{m \neq 0} C_{m,0} \left( 2|a_{p+m,y}|^2 + |a_{p+m,x}|^2 \right) a_{p,y}.
\end{align}
These terms can be more accurately modeled using phase rotation \cite{Luo_2023}:
\begin{align}\label{PBC-DBP}
    a_{p,x} \exp\left( i\sum_{m\neq 0} C_{m,0} \left( 2|a_{p+m,x}|^2 + |a_{p+m,y}|^2 \right) \right) a_{p,x} \nonumber \\ 
    a_{p,y} \exp\left( i\sum_{m\neq 0} C_{m,0} \left( 2|a_{p+m,y}|^2 + |a_{p+m,x}|^2 \right) \right) a_{p,y}.
\end{align}

As we can see, \eqref{PBC-DBP} is equivalent to the nonlinear step in FDBP. The required N-filter is, in fact, equivalent to \( N_{xx} = N_{yy} = 2N_{xy} = 2N_{yx} = 2[C_{-m,0}, C_{-m+1,0}, \dots, C_{m-1,0}, C_{m,0}] \). With Gaussian pulse assumption, the PBC coefficient related to intral-channel XPM is real numbers which can be expressed as \cite{tao2011multiplier}:
\begin{equation} \label{iXPM_coeff}
    C_{m, 0}= \frac{8}{9} \frac{\gamma \tau^2}{\sqrt{3}\left|\beta_2\right|} \frac{1}{2} E_1\left(\frac{m^2 T^2 \tau^2}{3\left|\beta_2\right|^2 L^2}\right),
\end{equation}
where $L$ is transmission distance, $T$ is symbol time (also inverse of the symbol rate), $\tau$ is pulse width, $\beta_2$ is dispersion coefficient, $E_1$ is exponential integral function. 
We find that this relationship reveals the connection between N-filters at different symbol rates.
Since pulse width $\tau \approx T = \frac{1}{R_s}$, 
$$
C_{m,0} \approx T^2 f(T^2 m), \ f(x) = \frac{8}{9} \frac{i\gamma}{\sqrt{3}\left|\beta_2\right|} \frac{1}{2} E_1\left(\frac{x^2}{3\left|\beta_2\right|^2 L^2}\right).
$$
From this expression, we find that \( C_{m,0} \) is a symmetric real kernel which fundamentally dependent only on the transmission baud rate \( R_s = \frac{1}{T} \), and is independent of the WDM channel count \( N_{ch} \) and signal power \( P_{ch} \). Moreover, when $R_s$ changes, \( C_{m,0} \) can be interpreted as a scaling transformation of the same functional form, with the scaling factor being \( T^2 \). 

\begin{figure}
    \centering
    \includegraphics[width=0.45\textwidth]{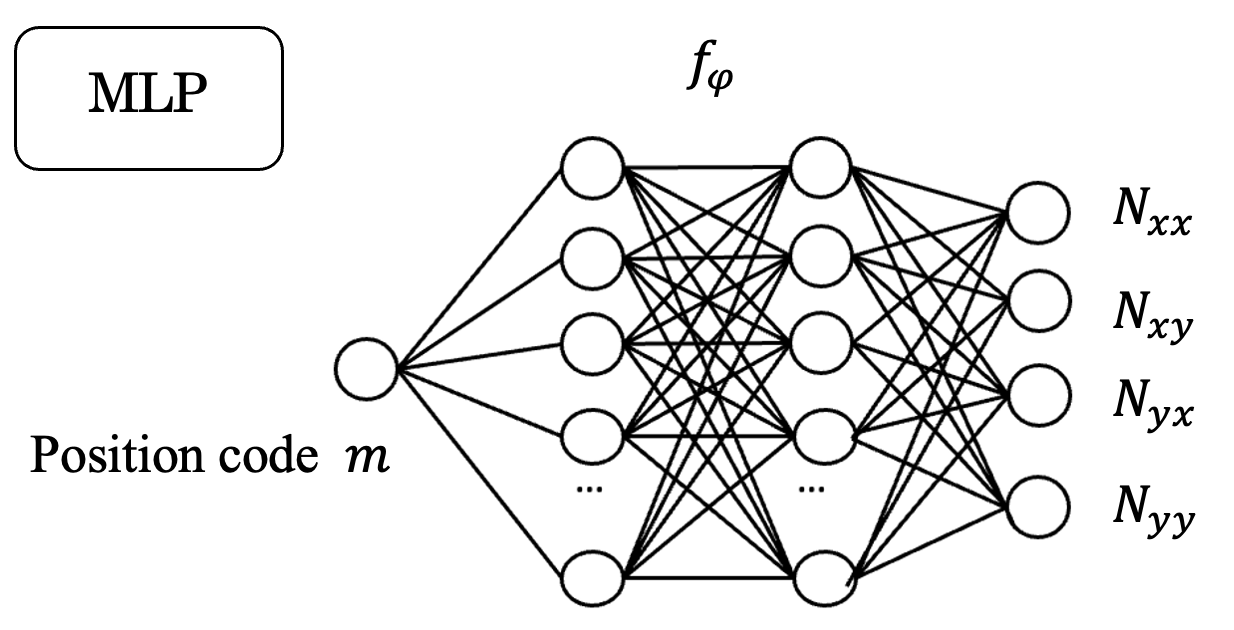}
    \caption{The N-filter generated by the hyper-network $f_{\varphi}$ for single-polarizarion or dual-polarization signals.}
    \label{MLP}
\end{figure}

In fact, the waveform we use is not exactly Gaussian pulse, and in practical applications, many parameters in the system cannot be accurately estimated. Additionally, there are other amplifier noises and nonlinear effects in the system. Therefore, the expression of $f(x)$ is not precise. Consequently, we choose to use a neural network $f_{\varphi}$ to learn this function. For single-polarization signals, $f_{\varphi}$ is a function mapping from $R^1$ to $R^1$. For dual-polarization signals, $f_{\varphi}$ maps from $R^1$ to a $R^4$  output represent $[N_{xx}, N_{xy}, N_{yx}, N_{yy}]$, as shown in the Fig \ref{MLP}. The N-filter is given as:
\begin{equation}\label{N-filter}
\mbox{N-filter} =T^2 f_{\varphi}(T^2 |m|).
\end{equation}
Note that in \ref{N-filter}, we employ symmetric positional encoding $|m|$ to apply $f_{\varphi}$ element-wise for generating the N-filter:
$$
|m| = \left[\frac{N_f-1}{2}, \frac{N_f-3}{2}, \ldots, -1, 0, 1, \ldots, \frac{N_f-1}{2}\right].
$$
This naturally introduces symmetry into the N-filter and effectively reduces the number of model parameters.
% From here, we can distinguish the difference between FDBP and Meta-DBP. 
% The N-kernel of FDBP is a fixed parameter that does not change once training is completed. 
% In contrast, the N-kernel of Meta-DBP is the output of the hypernetwork $f_{\varphi}$ and varies according to the different task-specific information it receives. 
% Our simulations in Section 3 validate that the design of Meta-DBP can generalize better across various modes of fiber data compared to the original FDBP.

In the Meta-DBP layer, see Fig \ref{Meta-DBP}, the input includes not only the signal $u_j$ but also a task-specific information $e=(R_s, N_f)$ including symbol rate $R_s$ and the length of the nonlinear filter $N_f$. Despite our dataset simulating different WDM channel counts $N_{ch}$, different signal powers $P_{ch}$, and different transmission baud rates $R_s$, our comparative simulations revealed that FDBP generalizes well over $N_{ch}$ and $P_{ch}$ but poorly over $R_s$. Therefore, we chose the symbol rate $R_s$ as the task-specific information. Additionally, due to the significant impact of the nonlinear filter length $N_f$ on the model's performance \cite{fan2021combined} \cite{CNNDBP}, we also included $N_f$ as task-specific information for the model. The optimal value of $N_f$ is determined by increasing its value until the performance converges, balancing performance gains and computational efficiency (in our experiments, we set $N_f = 401$ for DBP step=5, $N_f=201$ for DBP step=25).
% More simulation results for $N_f$ are further detailed in section 3. 
The algorithm of Meta-DBP can be roughly divided into two steps. The first step is to generate the D-filter and N-filter using the task-specific information $e$, which are used to compensate for dispersion and nonlinear phase rotation, respectively. The second step is to perform dispersion compensation and nonlinear compensation using the D-filter and N-filter, respectively, as shown in Figure \ref{Meta-DBP}. The generation of the D-filter is done using Equation \ref{D-eq-time-dis}. The N-filter is generated by inputting the position encoding plus the scale transformation of the symbol time squared into a neural network, as shown in Equation \ref{N-filter}.

\subsection{XPM-ADF}
In this subsection, we present the methodology for selecting an appropriate adaptive filter to mitigate XPM. Previous studies have thoroughly investigated XPM-induced nonlinear interference \cite{dar2013properties,secondini2012analytical,dar2014inter,dar2017nonlinear}. Building upon these works, XPM noise can be effectively characterized as a time-varying Inter-Symbol Interference (ISI) process \cite{secondini2012analytical,dar2014time,2016VL}. The mathematical formulation of the XPM noise model is given by:
\begin{equation}\label{ch2:XPM_model}
\Delta a^{\text{XPM}}_{n} = \sum_{l} H_l^{(n)} a_{n+l}
\end{equation}
where $a_n = [a_{n,x}, a_{n,y}]^T$ denotes the transmitted symbol vector, and $H_l^{(n)}$ represents time-varying $2\times 2$ coefficient matrices. It is worth noting that accurate computation of $H_l^{(n)}$ requires complete knowledge of the interfering channel's transmitted symbols \cite{2016VL}, which is generally unavailable in practical systems. However, empirical studies have revealed that $H_l^{(n)}$ exhibits specific statistical properties \cite{dar2017nonlinear}, with the zeroth-order term $H_{0}^{(n)}$ dominating the XPM effect as phase noise.

To address the phase-dominant characteristic of XPM noise, we propose the XPM-ADF, which combines a FIR with a single-tap phase estimator. While architecturally similar to \cite{DDLMS}, our design specifically targets XPM compensation. The phase estimator rapidly tracks the dominant XPM phase noise ($l = 0$), while the FIR filter captures higher-order XPM effects ($l>0$) using oversampled data. FIR filter and phase estimator are jointly updated via the Normalized Least Mean Squares (NLMS) algorithm. The XPM-ADF operates as follows as shown in Fig \ref{MetaDSP}. First, the output signal is computed:
\begin{align*}
& \mathbf{u'}[n] = \mathbf{w}[n] \mathbf{u}_n= \left[\begin{array}{ll}
    \mathbf{w}_{x x}[n] & \mathbf{w}_{x y}[n] \\
    \mathbf{w}_{y x}[n] & \mathbf{w}_{y y}[n]
    \end{array}\right]\left[\begin{array}{l}
    \mathbf{u}_{x}[n] \\
    \mathbf{u}_{y}[n]
    \end{array}\right] \\
& \hat{\mathbf{y}}[n]  = \mathbf{v}[n]  \circ  \mathbf{u'}[n] = \left[\begin{array}{l}
    v_x[n] \cdot \mathbf{u'}_{x}[n] \\
    v_y[n] \cdot \mathbf{u'}_{y}[n]
    \end{array}\right], \\
& \mathbf{y}_D[n] = \mathcal{D}(\hat{\mathbf{y}}[n])
\end{align*}
where $\mathbf{w}[n] \in \mathbb{C}^{2\times 2\times d}$ is the FIR filter, $\mathbf{v}[n] \in \mathbb{C}^2$ is the phase estimator, $\mathbf{u}[n] \in \mathbb{C}^{2\times d}$ contains the input samples, $d$ is the filter length, $\circ$ is element-wise multiplication. $\mathcal{D}$ is the hard decision for 16QAM. Then, the phase estimator can be updated as follows:
\begin{align*}
& \mathbf{e}_v[n] = \mathbf{y}_D[n] - \hat{\mathbf{y}}[n] \\
& \mathbf{v}[n+1] = \mathbf{v}[n] + \frac{\mu_v}{|\mathbf{u'}[n]|^2 + \epsilon} \mathbf{e}_v[n] \mathbf{u'}[n]^*
\end{align*}
For the FIR filter, the error signal and update rule are:
\begin{align*}
&  \mathbf{e}_w[n] = \mathbf{y}_D[n] \frac{|\mathbf{v}[n]|}{v[n]} - \mathbf{u'}[n] \\
& \mathbf{w}[n+1] = \mathbf{w}[n] + \frac{\mu_w}{|\mathbf{u}[n]|^2 + \epsilon} \mathbf{e}_w[n] \mathbf{u}[n]^H
\end{align*}
where $\mu_v, \mu_w$ us the step size for phase estimator and FIR filter, $\epsilon$ are constants to avoid division by zero.

\subsection{Training phase}
\begin{figure*}[htbp]
    \centering
    \includegraphics[width=0.9\textwidth]{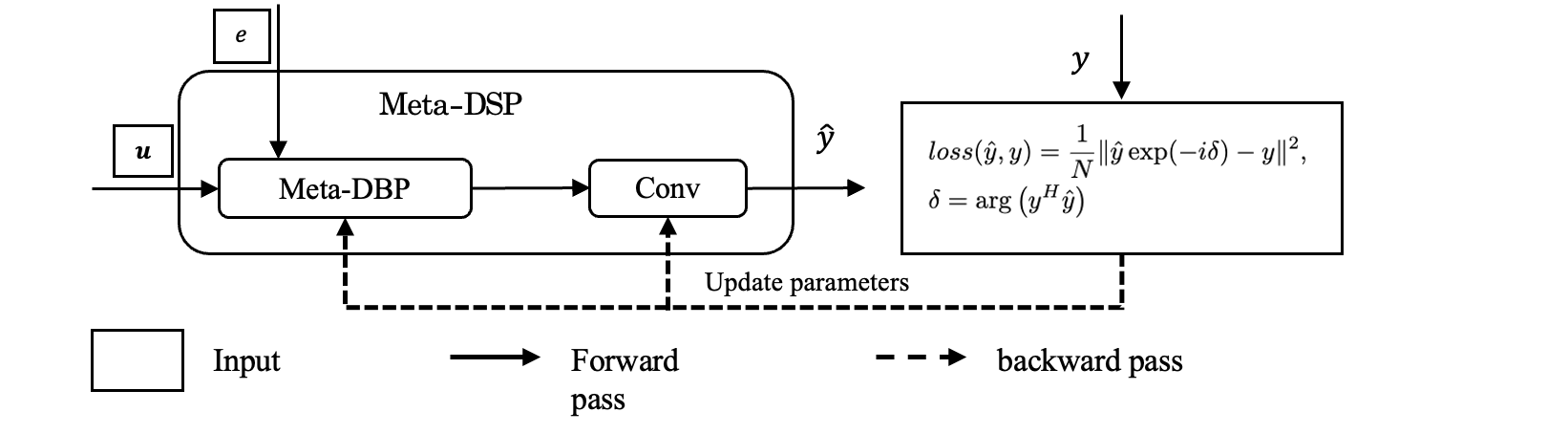}
    \caption{Schematic of the Meta-DSP training phase. In traning phase, we employs a 1D convolutional layer to substitute XPM-ADF and incorporates a custom loss function combining a average phase estimator with MSE, which we empirically demonstrate to yield more stable and faster convergence compared to end-to-end training. The hypernetwork $f_{\varphi}$ and convolutional layer are optimized via backpropagation with learning rates of 0.003 and 0.03, respectively.}
    \label{MetaDSP-train}
\end{figure*}

In this subsection, we present the training scheme for Meta-DBP. The structure of Meta-DSP includes two modules: Meta-DBP and XPM-ADF. The former is a highly parallelizable convolutional structure with many trainable parameters, while the latter is a recursive filter, a typical serial computation structure with no trainable parameters.
One option is to treat Meta-DBP and XPM-ADF as a single entity and perform end-to-end training, as done in \cite{fan2021combined}. However, in our attempts, we found that this training method is not stable. The training results are sensitive to hyperparameters such as batch size, and it is difficult to fully utilize the highly parallel computing capabilities of the GPU.
Ultimately, we chose to train the Meta-DBP module separately, as shown in Figure \ref{MetaDSP-train}. 
To ensure consistency between the model's performance during the training and testing phases, 
we used a static convolutional layer combined with an direct phase estimator to replace XPM-ADF. This approach is similar to the one proposed by \cite{hager2020}.

During the training stage, the input optical signal samples $\mathbf{u}$  are at a rate of 2 samples per symbol with shape $[B, 2W, N_{modes}]$.
$B$ is the batch size,
$W$ is the truncated signal length,
$N_{modes}$ is the number of polarization modes of the optical signal (1 or 2).
The task information $e$ has a shape of $[B, 2]$, where the two dimensions represent the symbol rate $R_s$ and the nonlinear filter length $N_f$.
The shape of the output signal from Meta-DBP is $[B, 2W - 2O, N_{modes}]$, where $2O$ is the length reduction caused by the valid convolution operation in Meta-DBP. At this stage, the signal still maintains a rate of 2 samples per symbol.
The static convolutional layer then downsamples the output signal from Meta-DBP to 1 sample per symbol, denoted as $\hat{y}$ with shape $[B, W-O, N_{modes}]$. The predicted signal $\hat{y}$ is then compared to the true signal $y$ from the transmitter for loss function calculation:
$$
\mbox{loss}(\hat{y}, y) =  \left|\hat{y} \exp(-i\delta) - y\right|^2, \delta = \angle \left(\hat{y}^H y\right).
$$
Where $\delta$ is the average phase error with shape $[B, 1, N_{modes}]$, the dot operation between $\hat{y}^H, y$ is carried on the time axis.
Finally, the loss function is utilized to perform backpropagation to update the parameters of the hypernetwork $f_{\varphi}$. 
During the testing phase, we should replace the static convolutional layer with the XPM-ADF and remove the direct phase estimator.

Finally, we outline the details of our training process. The hypernetwork MLP used a three hidden layers fully connected network with neurons in each layer set to [1, 200, 50, 2, 4], and ReLU was used as the activation function. The input dimension is 1, representing any element of the position code $|m|$. The output dimension is 4, corresponding to 4 N-filters in the dual-polarization case. Hidden layers are configured as 200, 50, and 2 units. The 200 and 50-unit layers enhance representation capacity, while the 2-unit layer exploits the symmetry of the 4 N-filters, allowing them to be represented by 2 parameters. Each dual polarization signal contained $5 \times 10^5$ symbols. During the training of FDBP and Meta-DSP, we used 40 Gbaud, 1-channel signals with four different energy levels. For testing, each mode used $1 \times 10^5$ symbols. We employed the Adam optimizer with a learning rate of 0.003 for Meta-DBP and learing rate 0.03 for Conv layer, a batch size of 10, and a signal truncation length of 2000, training for 20 epochs.

\subsection{Complexity analysis}
In this subsection, we investigate the computational complexity of all models. We adopt the real multiplications per symbol (RMPS) metric, a widely used measure in optical communication systems to quantify computational load during the design phase. We focus on several baseline models: CDC, DBP, and FDBP. To mitigate residual phase errors, a XPM-ADF is incorporated into each of these models. In the XPM-ADF, the filter function \(h\) is computed, and the gradient \(g_k\) is obtained. This process requires nearly \(2(\text{taps}+1)\) complex multiplications. Thus, the RMPS of XPM-ADF can be expressed as
\begin{equation}\label{CC_DDLMS}
C_{\text{XPM-ADF}} = 4\times 2 (\text{taps}+1),
\end{equation}
where the factor of 4 signifies that one complex multiplication can be represented through four real multiplications.

CDC stands out as a widely adopted commercial technology. It solely employs a filtering operation with a filter length \(N_d\), resulting in the lowest RMPS \cite{EDC_CC}:
\begin{equation}\label{EDC_CC_formula}
C_{\text{CDC}} = \min_{N} \frac{8N(\log_2(N) + 1)}{N - N_{d} + 1},
\end{equation}
This formula originates from the overlap and save method applied when utilizing FFT for convolution. The FFT size is denoted by \(N\), and we have optimized this size to minimize computational complexity.

For DBP, its complexity is largely influenced by the steps per span \(N_{stps}\). A higher \(N_{stps}\) value indicates improved compensation and increased RMPS. The DBP requires \(N_{span}N_{stps}\) convolution operations, where each filter has a length of approximately \(N_d/(N_{span}N_{stps})\). In addition to this, performing non-linear rotations also requires two complex multiplications (SpS = 2). Therefore, the RMPS for DBP can be articulated as:
\begin{equation}\label{DBP_CC_formula}
\begin{split}
C_{\text{DBP}} = \min_{N} 4N_{span}N_{stps}\left(\frac{2N(\log _2 N+1)}{N-N_d/(N_{span}N_{stps})+1}+2\right),
\end{split}
\end{equation}
where \(N_{span}\) denotes the total number of spans and \(N_{stps}\) indicates the propagation steps for each span.

The FDBP algorithm, outlined in section 2, is configured in our study with a back-propagation step count of 5 (\(N_{stps}=0.2, N_{span}=25\)), aligning with the non-linear kernel length \(N_f=401\) in the Meta-DSP model. The RMPS of FDBP adds a convolution cost with a filter of length \(N_f\) to the DBP method at each step:
\begin{equation}\label{FDBP_CC_formula}
\begin{split}
& C_{\text{FDBP}} = \min_{N} 4N_{span}N_{stps} \cdot  \\
& \left(\frac{2N(\log _2 N+1)}{N-N_d/(N_{span}N_{stps})+1} + \frac{2N(\log _2 N+1)}{N-N_f +1} \right),
\end{split}
\end{equation}

The architecture of Meta-DBP parallels that of the FDBP model. The only added computation stems from the hyper-network \(f_{\varphi}\). As this computation is executed only once and its load is distributed across all transmitted symbols, its impact is minimal. Thus, the RMPS of Meta-DSP matches that of FDBP:
\begin{equation}\label{Meta_DBP_formula}
\begin{split}
& C_{\text{Meta-DBP}} = \min_{N} 4N_{span}N_{stps}\cdot  \\
& \left(\frac{2N(\log _2 N+1)}{N-N_d/(N_{span}N_{stps})+1} + \frac{2N(\log _2 N+1)}{N-N_f +1} \right).
\end{split}
\end{equation}
\section{Transmission system model}
\begin{figure*}[htbp]
    \centering
    \includegraphics[width=0.98\textwidth]{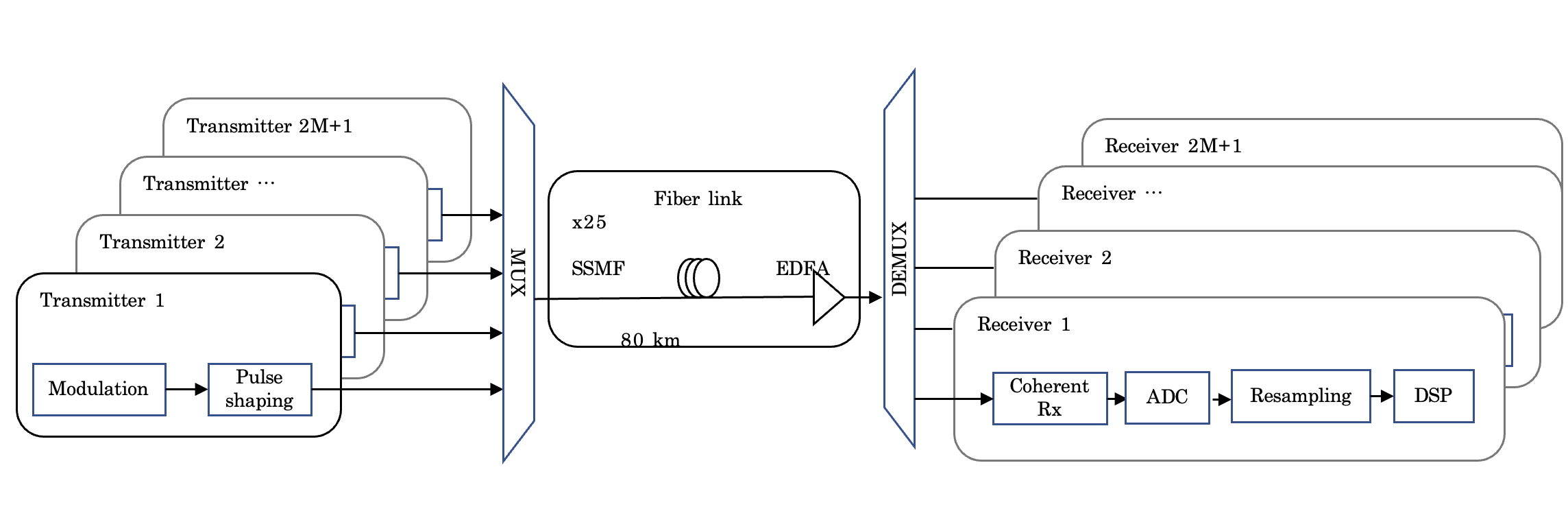}
    \caption{Scheme of simulated transmission link. $N_{ch}=2M+1$ WDM channels are multiplexed (i.e., the MUX module) in transmitter and transmission in fiber link. The transmission channel is composed of 25 segments of Standard Single-Mode Fiber (SSMF), each followed by an EDFA to amplify the optical signal, which simultaneously introduces ASE. At the receiving end, a demultiplexer (i.e., the DEMUX module) is used to split the optical signal into $2M+1$ channel signals, which then enter their respective coherent receivers for sampling. Finally, DSP is performed separately for each signal. }
    \label{Simulation model}
\end{figure*}

In this section, we introduce our simulated transmission system model. 
The simulated transmission link is depicted in Fig. \ref{Simulation model}. 
The transmitter multiplexes $N_{ch}=2M+1$ WDM channels, 
and the transmission channel consists of 25 spans of Standard Single-Mode Fiber (SSMF), 
each followed by an EDFA to amplify the optical signal, introducing amplified spontaneous emission (ASE). 
At the receiver end, a demultiplexer splits the optical signal into $2M+1$ channels, which then enter their respective coherent receivers for sampling. Finally, DSP is performed separately for each signal.

\subsection{Transmitter and Optical Fiber Link}
We studied a point-to-point WDM system employing dual-polarization 16QAM signals modulated with a raised-cosine function (roll-off factor = 0.1). To comprehensively evaluate the model's generalization capability across diverse scenarios, we generated multiple transmission configurations by varying three key parameters: WDM channel counts ($N_{ch}$) of 1, 3, 5, and 21 channels to examine XPM noise intensity effects; baud rates ($R_s$) of 40GBaud, 80GBaud, and 160GBaud, covering current and future high-speed systems; and channel power levels ($P_{ch}$) ranging from -8 to 8 dBm to assess nonlinearity impacts. The channel spacing was fixed at 0.2 times the symbol rate with a central wavelength of $\lambda$ = 1550 nm. The transmitted signal can be described as:
\begin{equation}\label{WDM time domain}
u(0,t) = \sum_{k=-M}^{M} A_k(0,t) \exp\left(i (\omega_k - \omega_0) t\right),
\end{equation}
where $\omega_k$ represents the frequency of the $k$th channel. Each channel's symbols are represented as:
\begin{equation}\label{Each channel symbols}
A_k(0,t) = \sqrt{P_k}\sum_{n=1}^{N} x^k_n g(t - nT),
\end{equation}
where $P_k=P_{ch}$ for all k denotes the launch power of the $k$th channel, N is number of symbols, and $x_n^k \in \mathcal{X}$ is a complex symbol in the IQ plane, with $\mathcal{X}$ representing the 16QAM modulation format constellation. The pulse shape function $g(t)$ is typically a raised-cosine function with a roll-off parameter of 0.1. For dual-polarization signals, $u$ and $A_k$ become vector signals with two dimensions representing the x and y polarization directions, retaining the same form as equations \eqref{WDM time domain} and \eqref{Each channel symbols}.

The generated signal is then transmitted through a simulated fiber link composed of 25 spans of 80km single-mode fiber, totaling 2000km. A standard EDFA with a 4.5 dB noise figure compensates for the losses of each span. Signal propagation is modeled by the NLSE in the single-polarization system and by the Manakov equations in the dual-polarization system. These equations are solved using a standard second-order symmetrical split-step Fourier method. All system parameters are summarized in Table \ref{Transmission parameters}.
\begin{table}[ht]
    \centering
    \begin{tabular}{cc}
        \hline \text{Parameter} & \text{Value} \\
        \hline \text{Attenuation} & 0.2 dB/km \\
        \text{Dispersion} & 16.5 ps/nm/km \\
        \text{Nonlinearity} & 1.31 /W/km \\
        \text{Distance} & \(25 \times 80\) km \\
        \text{Noise Figure} & 4.5 dB \\
        \text{modulation format} & 16QAM \\
        \text { COI Wavelength } & 1550 $\mathrm{~nm}$ \\
        \text { RRC Roll-off } & 0.1 \\
        \text { Symbol Rate } & 160/80/40 $\mathrm{GBaud}$ \\
        \text { number of channels } & 21/11/9/7/5/3/1  \\
         \text{polarization} & \text{single or dual} \\
        \hline
    \end{tabular}
    \caption{Transmission model parameters}
    \label{Transmission parameters}
\end{table}

\subsection{Receiver and DSP Model}
The model integrates a total of $N_{ch}$ modulated optical signals at the transmitter end, which are then propagated through the optical fiber. At the receiver end, these signals are de-multiplexed into $N_{ch}$ individual channels. Each channel signal is independently received and sampled at 2 samples per symbol. The sampled electrical signals then undergo DSP, the DSP workflow consists of two main stages: DBP and ADF. DBP compensates for both linear and nonlinear effects by solving the $z$-reversed NLSE in the digital domain, while the adaptive filter corrects the phase error caused by XPM and downsamples the signal to 1 sample per symbol.

Since the focus of this paper is on compensating for nonlinear noise in optical communication systems, we make the following simplifications when simulating the receiver: we do not introduce additional receiver noise and frequency offset, thus eliminating the need for a frequency offset estimator (FOE) and carrier phase estimator (CPE). All compensation simulations are conducted on the central channel, ensuring natural alignment of transmitted and received symbols, and obviating the need for a clock recovery module.

For comparison purposes, we also consider nonlinear equalizers based on the DBP method. 
After downsampling to SpS=2, Split-Step Fourier Method (SSFM) based DBP is applied to the central channel for SPM compensation. It should be noted that for the DBP method, we numerically optimized the nonlinear parameter, as its value depends on the dispersion coefficient, the number of propagation steps, and the launch power. Next, we compensate for the remaining nonlinear phase shift of all symbols using XPM-ADF. After nonlinear equalization, we apply demodulation and calculate the bit error rate (BER) for the central channel of interest (COI).
\section{Numerical Results}
In this section, we validate the effectiveness of Meta-DSP through numerical results. 
First, we explore the performance of XPM-ADF, showing that it captures XPM noise more effectively than other adaptive filters. Consequently, XPM-ADF is used in all subsequent simulations.
Next, we conduct a series of simulations to test the generalization performance of FDBP. Our results demonstrate that the FDBP+XPM-ADF structure generalizes well across two system parameters: the number of WDM channels ($N_{ch}$) and the power per channel ($P_{ch}$). However, we find that FDBP struggles to generalize when the transmission baud rate ($R_s$) changes, which motivates our Meta-DBP design.
Finally, we present numerical simulations that highlight the superior generalization performance of Meta-DSP.

\subsection{Comparison simulations of XPM-ADF}
This subsection evaluates the XPM compensation performance of six distinct filtering approaches: BPS, DD-LMS, DD-RLS, V\&V, our proposed XPM-ADF, and a static filter as baseline. We conducted numerical simulations on dual-polarization 80 Gbaud systems with 21 channels at 2 dBm launch power (near system optimum). The optical signals first undergo digital backpropagation (DBP) with 64 steps per span for SPM compensation, followed by XPM compensation using the selected filter module.

All algorithms were carefully optimized through detailed parameter tuning. For BPS~\cite{BPS}, we found the best performance with a 40-symbol window length and 161 phase candidates, as further increases did not improve the Q-factor. The V\&V algorithm, implemented as in~\cite{fatadin2010laser}, worked best with a 121-symbol window length. Using a thorough grid search, we identified the optimal settings for other methods: DD-LMS performed best with a 32-tap filter and step size $\mu=3\times10^{-4}$, while DD-RLS used an 8-tap filter (to limit complexity) with a forgetting factor of 0.996 and $\delta=0.01$. Our XPM-ADF used a 32-tap filter with $\mu_v=\mu_w=2^{-6}$, and a static 32-tap filter served as a reference for dynamic noise.

\begin{figure}[ht]
    \centering
    \includegraphics[width=0.52\textwidth]{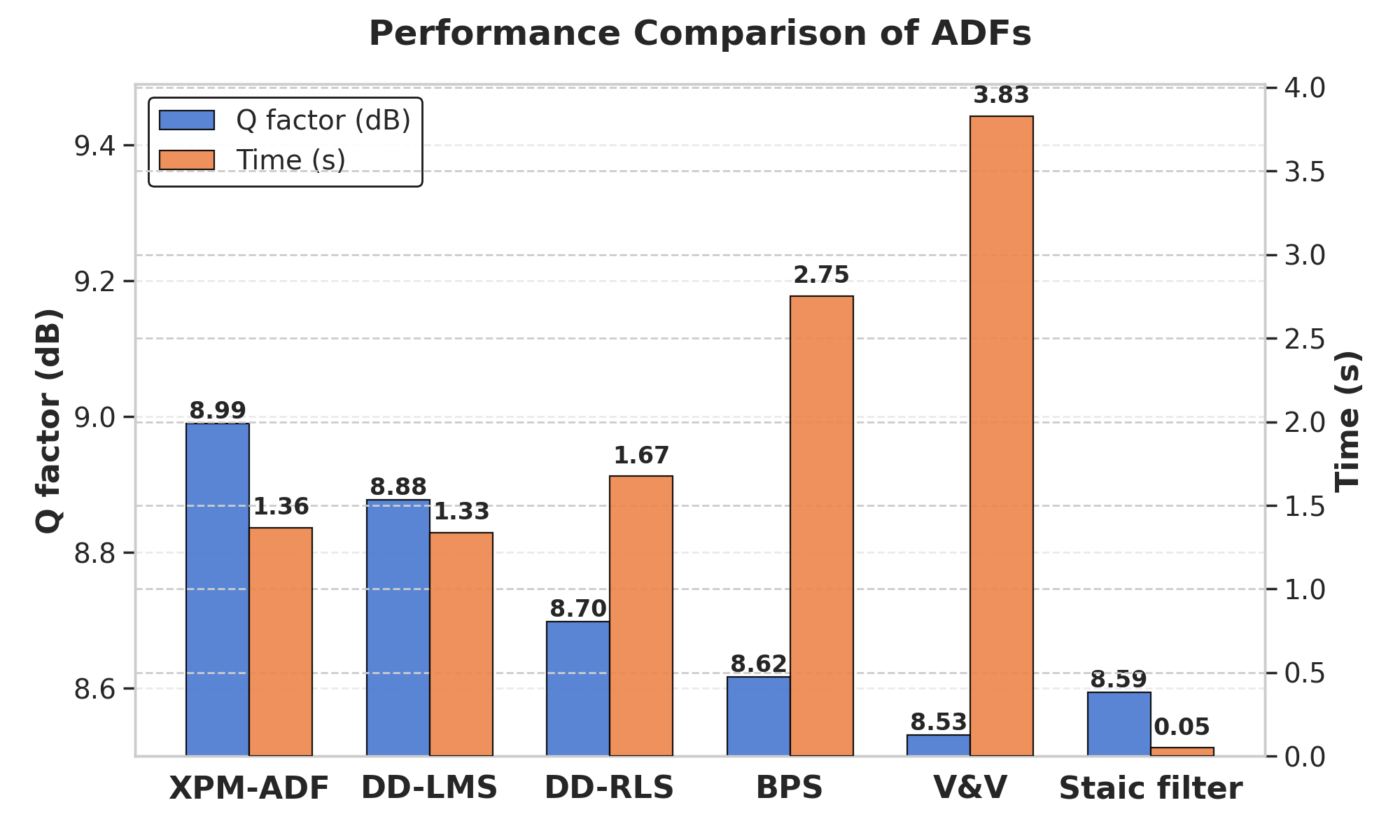}
    \caption{Performance comparison of various compensation methods (80 Gbaud, 21 channels, dual-polarization). XPM-ADF demonstrates optimal balance between Q-factor gain and computational cost.}
    \label{fig:all_filters}
\end{figure}

Figure~\ref{fig:all_filters} compares Q-factor and runtime of five algorithms implemented on a single CPU. Among them, XPM-ADF achieves the highest Q-factor improvement while maintaining relatively low computational latency, demonstrating its superior efficiency. DD-LMS and DD-RLS follow in performance, though DD-RLS suffers from computational complexity constraints that limit its filter length, resulting in inferior XPM compensation capability compared to DD-LMS. The remaining algorithms—Static Filter, BPS, and V\&V—exhibit notably poor performance. The Static Filter's weak results indicate strong dynamic noise in the signal, while the inadequacy of BPS and V\&V suggests that phase-only modeling is insufficient for effective XPM noise compensation.

\begin{figure*}[htbp]
    \centering
    \includegraphics[width=0.95\textwidth]{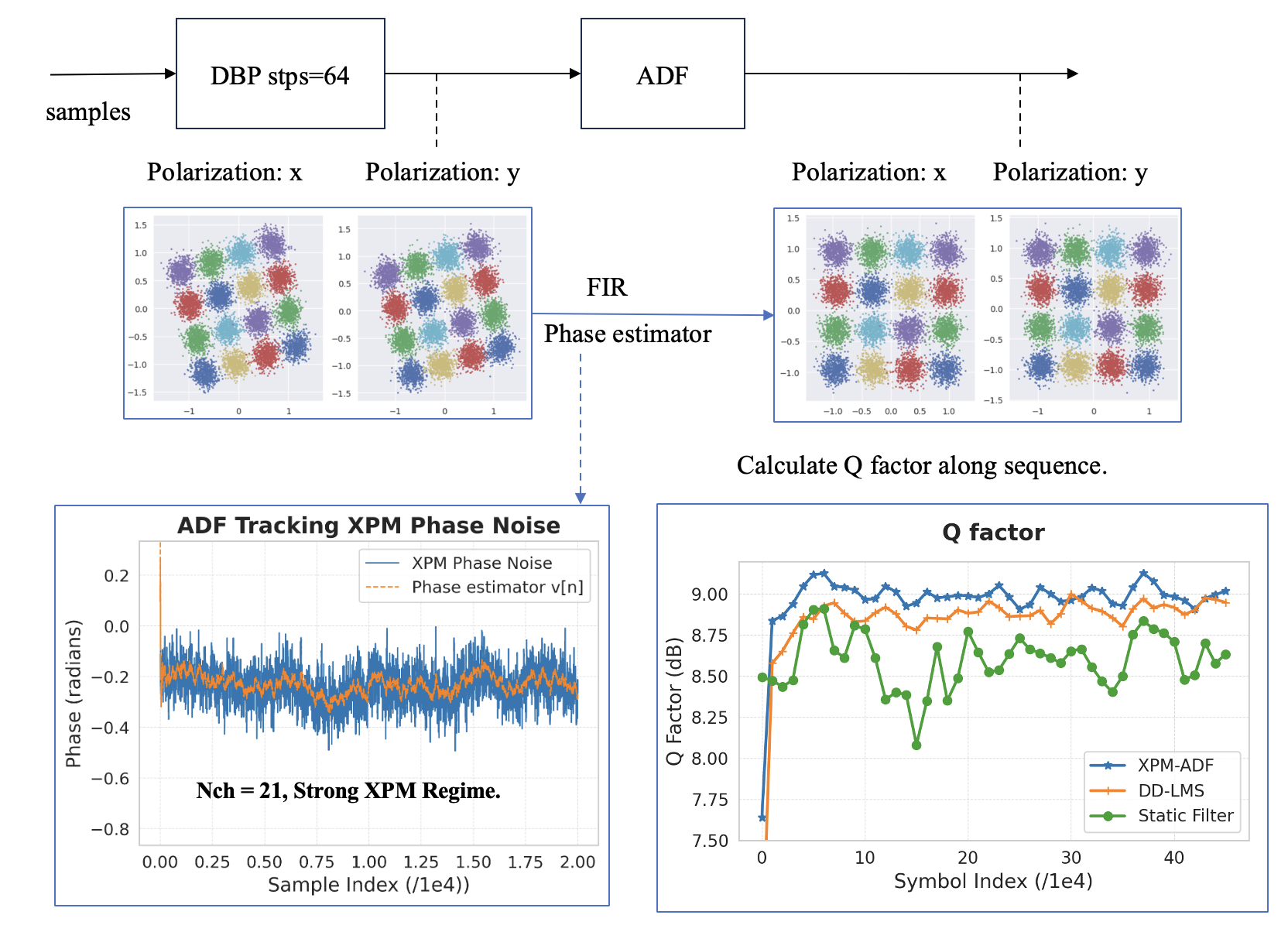}
    
    \caption{XPM compensation (80 Gbaud, 21 channels, dual polarization).
    The upper plot shows the DSP pipeline. We design a DBP module with 64 steps per span to compensate for SPM, followed by different adaptive filters to compensate for XPM. The lower left plot shows the XPM phase noise variation over time and the phase estimator $\mathbf{v}[n]$ tracking results of XPM-ADF under WDM 21 channels. By plotting the XPM phase noise and the phase estssimator of XPM-ADF, we can observe that the XPM noise can be effectively tracked by the phase estimator of XPM-ADF under WDM 21 channels. The lower right plot compares the performance of BPS, DD-LMS, XPM-ADF, static filter in terms of Q factor. It is evident that XPM-ADF outperforms the other algorithms in capturing XPM noise.} 
    \label{test_ADF}
\end{figure*}

Figure~\ref{test_ADF} demonstrates that DBP corrects the optical signal to a regular shape, leaving only residual phase errors caused by XPM noise. By comparing DBP-processed symbols with the transmitted symbols, we extract the XPM noise, as illustrated in the lower left plot. The XPM phase noise exhibits strong temporal correlation with slow dynamics, making it trackable by the ADF. The lower left plot of Figure~\ref{test_ADF} further shows that the phase estimator $v[n]$ successfully tracks the XPM phase noise over time, enabling XPM-ADF to outperform other algorithms in terms of Q factor. 

\begin{figure*}[htbp]
    \centering
    % First image
    \includegraphics[width=0.95\textwidth]{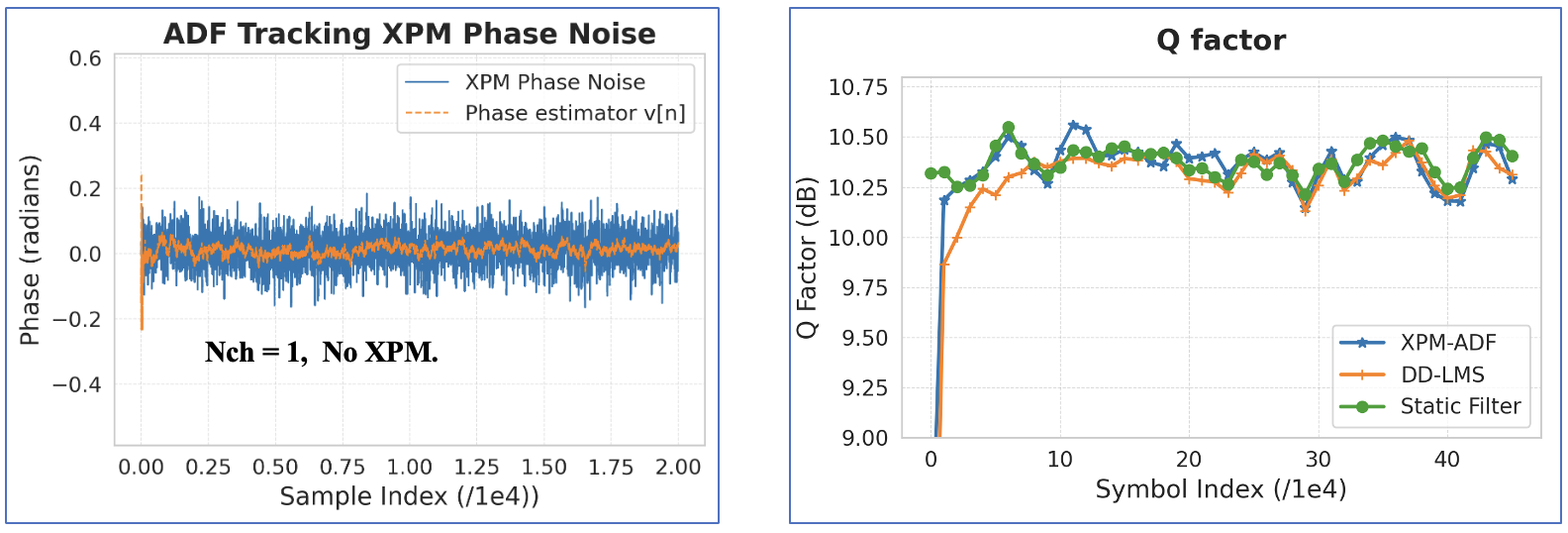}
    \caption{XPM compensation (80 Gbaud, 1 channel, dual polarization). In the single-channel signal case, without XPM noise, the phase tracker in XPM-ADF shows almost no dynamic change over time. However, in the case of 21 channels, the phase noise tracked by XPM-ADF becomes clearly noticeable. The Q-factor results for XPM-ADF, DD-LMS, and static filter are almost identical in the single-channel case, but show significant differences in the 21-channel case (Fig \ref{test_ADF}). This indicates that XPM-ADF is effective in capturing XPM noise.}
    \label{XPM_compare}
\end{figure*}

To validate the effectiveness of XPM-ADF in capturing XPM noise, we conducted additional tests using a single-channel dataset, where XPM noise is inherently absent. The results of this evaluation are presented in Figure~\ref{XPM_compare}. In the single-channel scenario, the XPM noise level remains negligible, indicating minimal interference. In contrast, the phase noise observed in the 21-channel case (as shown in Fig.~\ref{test_ADF}) exhibits significant random fluctuations over time, highlighting the presence of XPM-induced disturbances. The Q-factor analysis further supports these findings. In the single-channel case, the performance of XPM-ADF, DD-LMS, and the Static filter algorithms is nearly identical, as expected due to the absence of XPM noise. However, in the 21-channel scenario (Fig.~\ref{XPM_compare}), the Q-factor achieved by XPM-ADF is significantly higher than that of the other two algorithms. This improvement confirms the ability of XPM-ADF to effectively capture and mitigate XPM noise, thereby enhancing system performance in multi-channel environments.

\subsection{Transferability of the FDBP Model}
In this subsection, we test the transferability of the FDBP model on multimodal optical fiber data by revisiting our data generation process. 
The optical fiber signal data has three important system parameters: the number of WDM channels $N_{ch}$, 
the symbol rate $R_s$, and the average power per channel $P_{ch}$. 
To test the model's ability to transfer across different modes, we designed three sets of simulations using a controlled variable approach. 
In each setting, we fix two parameters and vary the other to create different training datasets for comparison. The performance of models trained on different training sets is then tested on a unified test set as shown in Table \ref{table:transferability}.
\begin{table*}[ht]
\centering
\begin{tabular}{|c|c|c|}
\hline
Simulation & Training Set & Test Set \\
\hline
1 & Dual polarization 16QAM. Rs=80 Gbaud, Nch=21, Pch=-2 or 2 or 6 & Dual polarization 16QAM. Rs=80 Gbaud, Nch=21, Pch=-3,-2,\ldots,6 \\
\hline
2 & Dual polarization 16QAM. Rs=80 Gbaud, Pch=[-2,0,2,4], Nch=1 or 5 or 21 & Dual polarization 16QAM. Rs=80 Gbaud, Nch=21, Pch=-3,-2,\ldots,6 \\
\hline
3 & Dual polarization 16QAM. Nch=21, Pch=[-2,0,2,4], Rs=40,80, 160 Gbaud & Dual polarization 16QAM. Rs=80 Gbaud, Nch=21, Pch=-3,-2,\ldots,6 \\
\hline
\end{tabular}
\caption{Simulation setup for testing the transferability of the FDBP model.}
\label{table:transferability}
\end{table*}

\begin{figure}[htbp]
    \centering
    \includegraphics[width=0.4\textwidth]{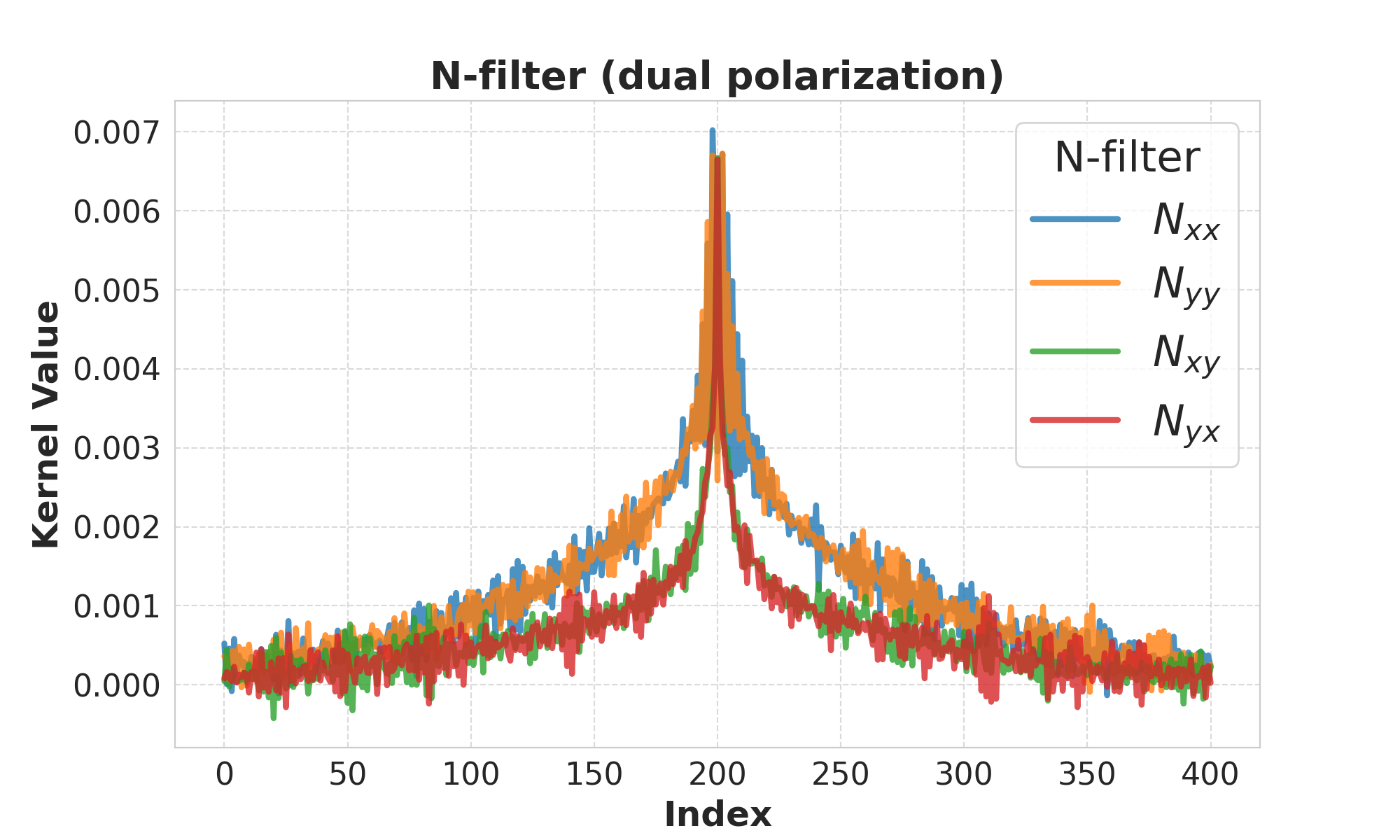}
    \includegraphics[width=0.4\textwidth]{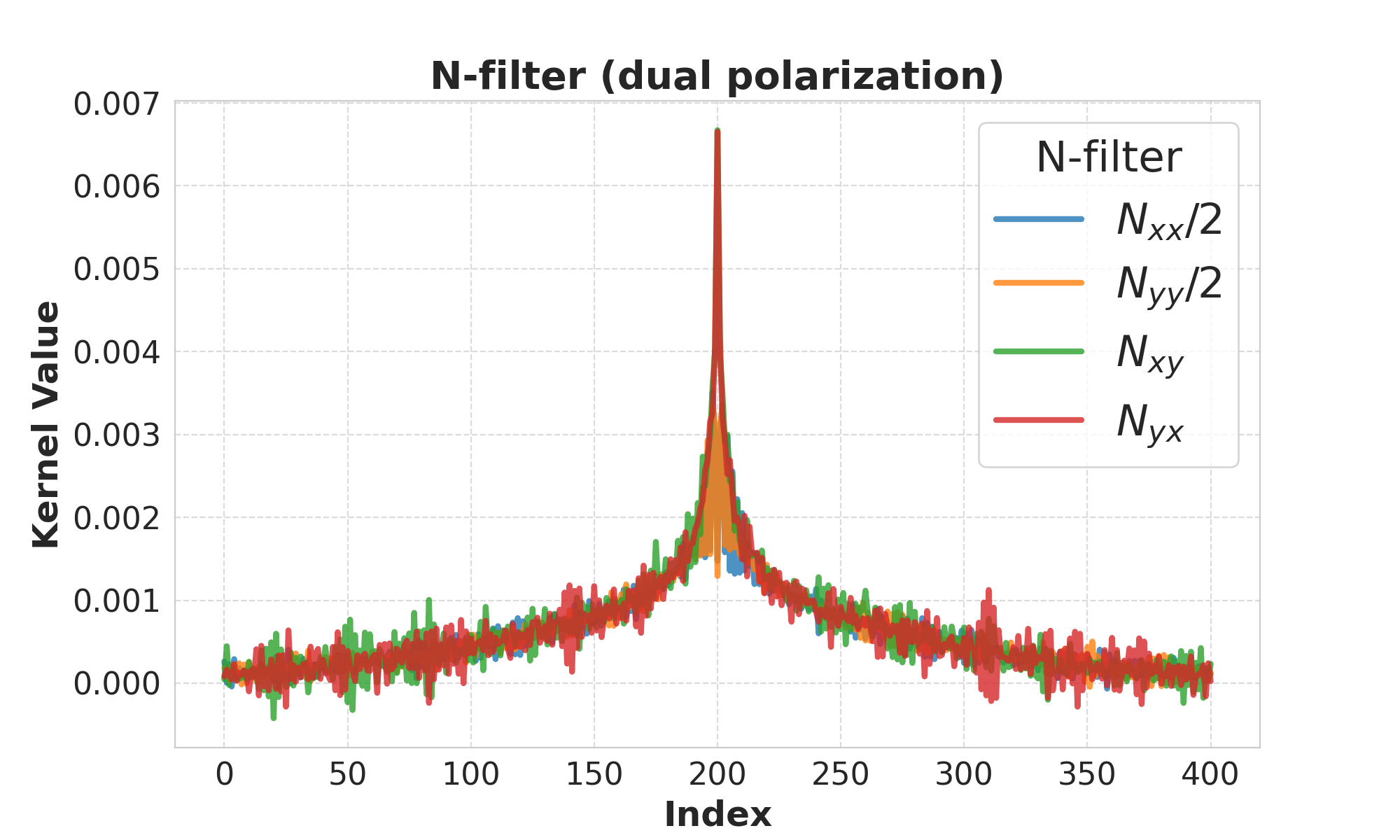}
    \caption{Learned N-filters for dual polarization. The figure above shows the four results of the N-filter for dual polarization: \( N_{xx}, N_{xy}, N_{yx}, N_{yy} \). The figure below shows the values of \( N_{xx} \) and \( N_{yy} \) divided by 2, revealing significant overlap. This leads to the conclusion that \( N_{xx} = N_{yy} = 2N_{xy} = 2N_{yx} \).}
    \label{fig:N_filter}
\end{figure}

\begin{figure}[htbp]
    \centering
    \begin{minipage}[b]{0.45\textwidth}
        \centering
        \includegraphics[width=\textwidth]{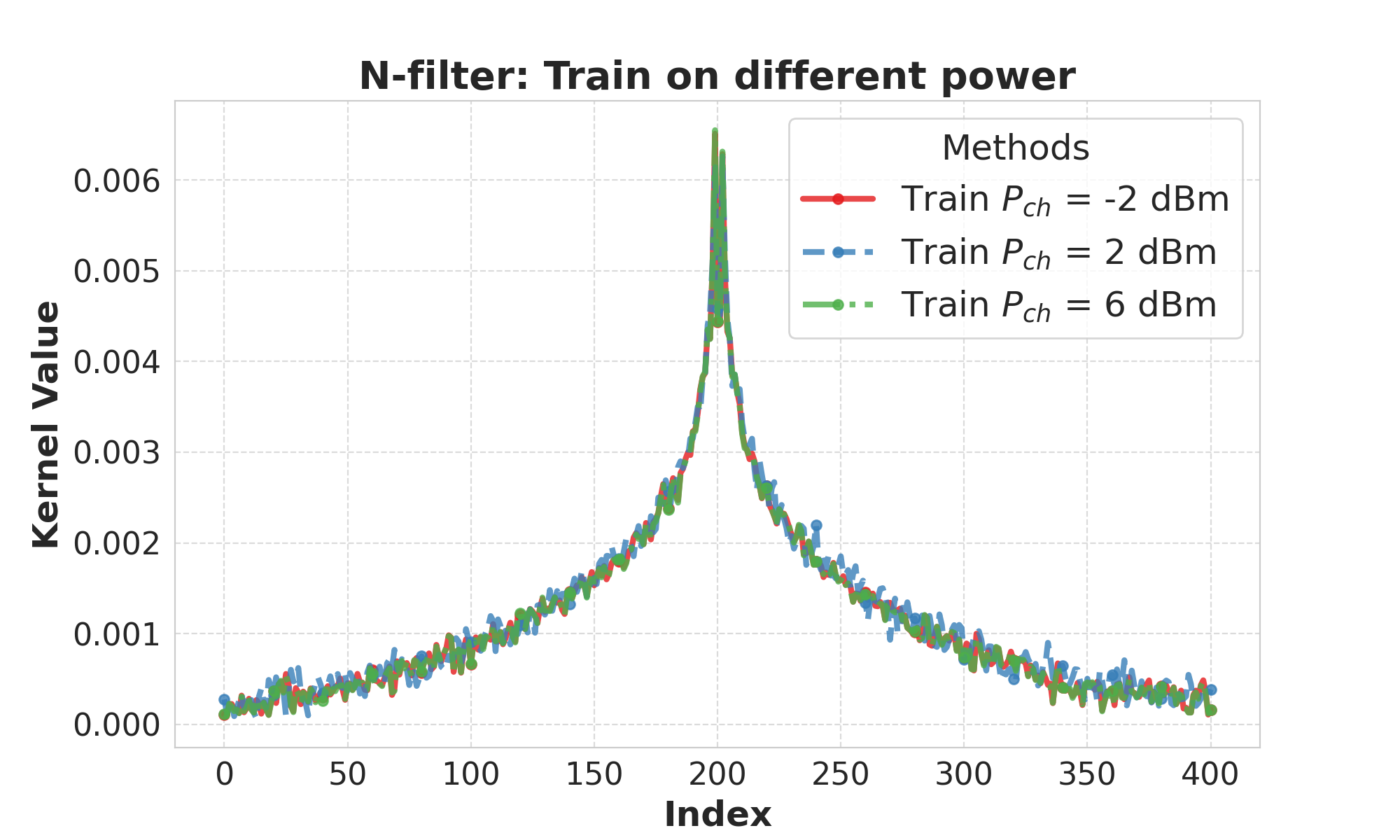}
        \caption*{(a) Simulation 1: FDBP's N-filters trained on different power $P_{ch}$}
    \end{minipage}
    \hspace{0.04\textwidth}
    \begin{minipage}[b]{0.45\textwidth}
        \centering
        \includegraphics[width=\textwidth]{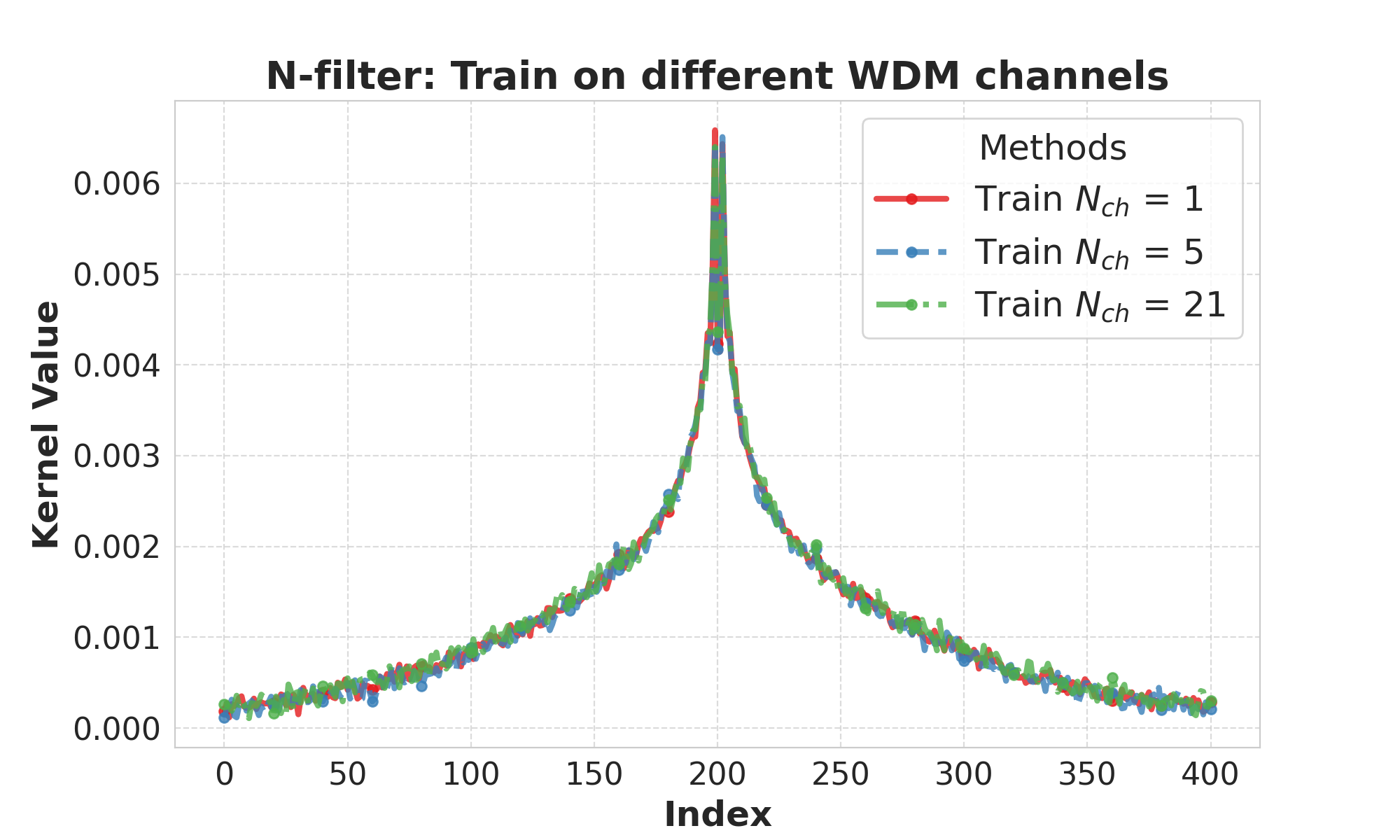}
        \caption*{(b) Simulation 2: FDBP's N-filters trained on different WDM channels $N_{ch}$}
    \end{minipage}
    
    \vspace{0.4cm}
    
    \begin{minipage}[b]{0.45\textwidth}
        \centering
        \includegraphics[width=0.96\textwidth]{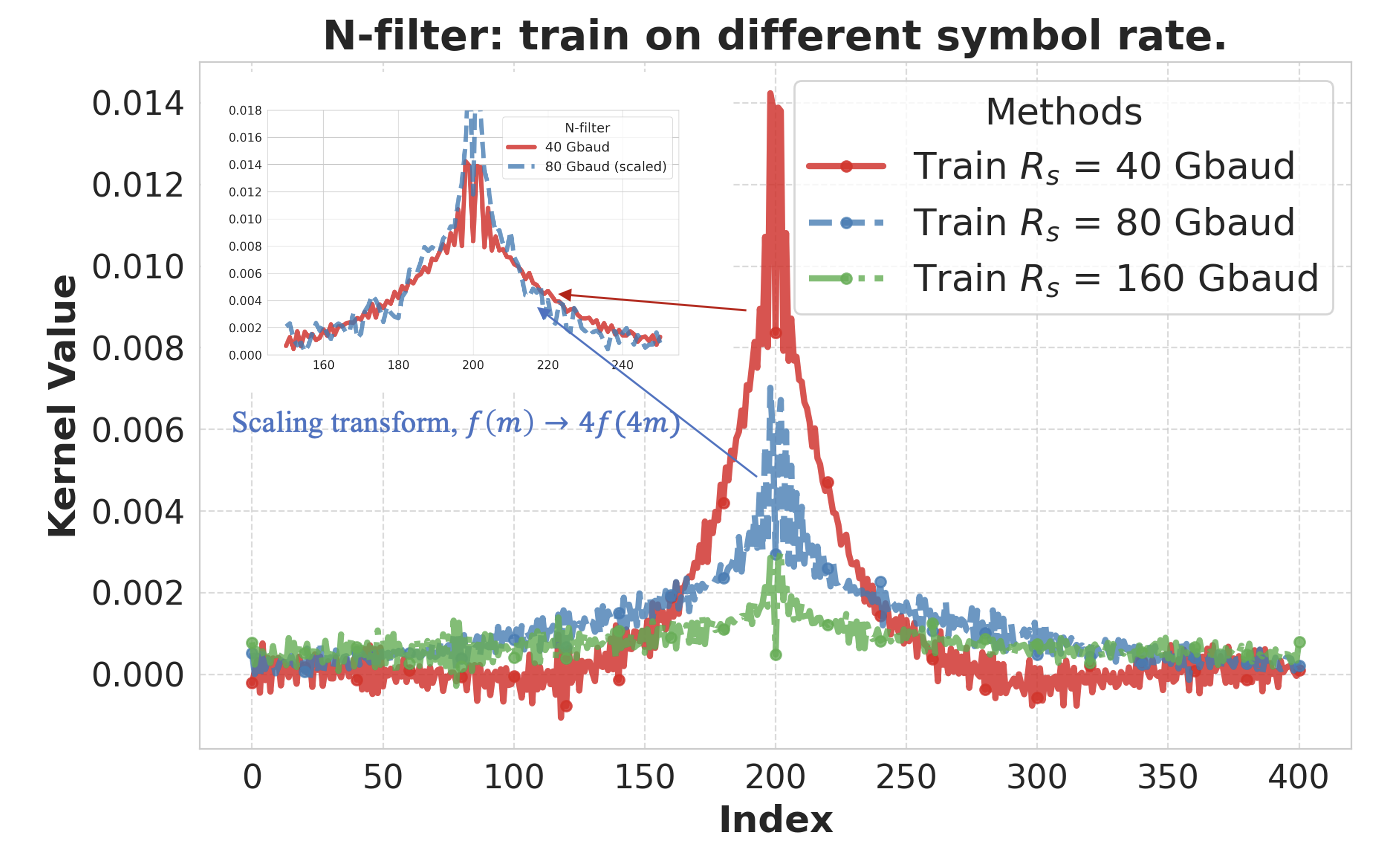}
        \caption*{(c) Simulation 3: FDBP's N-filters trained on different symbol rates $R_s$}
    \end{minipage}

    \caption{FDBP N-filter comparison for different system parameters: 
    The N-filter can be transferred across different $N_{ch}$ and $P_{ch}$ values, but a scaling transformation is needed for different $R_s$. For example, by compressing the N-filter learned at 80 Gbaud along the x-axis by 4 times and stretching it along the y-axis by 4 times, it closely matches the N-filter learned at 40 Gbaud.}
    \label{ex1-3}
\end{figure}

We first present a set of dual-polarization N-filter results in Fig. \ref{fig:N_filter}. The first plot shows the N-filter learned at 80 Gbaud with single channel. Here, $N_{xx}$, $N_{xy}$, $N_{yx}$, and $N_{yy}$ represent 4 N-filters of the interaction between the two polarization states. From the figure, it can be observed that the data-driven FDBP learns an approximate relationship: $N_{xx} = N_{yy} = 2N_{xy} = 2N_{yx}$. This result aligns with the theoretical analysis presented in Section 2.B. For simplicity, in the following figures comparing the three simulations, we will only show the results for $N_{xx}$.

In Fig. \ref{ex1-3}, we present N-filter ($N_{xx}$) for Simulations 1 to 3. The data reveal that the N-filter demonstrates strong consistency with variations in $N_{ch}$ and $P_{ch}$, but exhibits distinct shapes when the transmission rate $R_{s}$ changes. In the bottom panel of Fig. \ref{ex1-3}, we observe that the N-filters for different transmission rates appear to follow the same functional form, differing only by a scale factor. This observation aligns with the analysis presented in Section 2.B, which underpins the core concept of Meta-DBP. We will now discuss the reasons behind the results of these three simulations.

First, $N_{ch}$ only influences the intensity of XPM noise, whereas the primary goal of FDBP is to capture the SPM noise within the channel. As a result, FDBP models trained on data with different $N_{ch}$ values are essentially similar. Second, the form of FDBP is derived from the NLSE and Manakov equations, and its design inherently generalizes the $P_{ch}$ parameter. Finally, different transmission rates fundamentally affect the temporal statistical properties of SPM noise. Higher rates introduce longer time-memory effects of SPM at the symbol scale, which is why models trained at different rates in simulation 3 are not compatible.

\subsection{Meta-DSP Generalization Performance}
Based on the analysis from the previous subsection on the FDBP comparison numerical simulations, 
we chose the task inputs for Meta-DBP as $e=(R_s, N_f)$. 
The parameters $N_{ch}$ and $P_{ch}$ were not included in this input. 
This design allows Meta-DBP to naturally inherit the generalization performance of FDBP in terms of $N_{ch}$ and $P_{ch}$. Additionally, 
since we encoded the scaling relationship of N-filters under different transmission rates into the network structure itself, 
the final Meta-DSP can achieve excellent generalization performance across all modes. 
To train Meta-DSP effectively, we only need to collect a dataset of low-rate, single-channel, low-power data. Once the model is trained, it can be directly applied to any scenario including high-rate, multi-channel optical fiber signals.

In this subsection, we conduct numerical experiments to evaluate the generalization capability of Meta-DSP. Specifically, we train Meta-DSP models using a dataset configured with a symbol rate of 40 Gbaud and a single-channel setup. Subsequently, we test their Q-factor performance across a wide range of scenarios: symbol rates ranging from 40 Gbaud to 160 Gbaud and channel counts varying from 1 to 21. To assess the impact of model complexity, we implement two variants of Meta-DSP with different capacity levels (Step=5 and Step=25). For benchmarking purposes, we also train and tests FDBP under the same configuration. Furthermore, to examine the extent to which Meta-DSP achieves nonlinear compensation at 160 Gbaud, we additionally train an FDBP model with Step=25 using 160 Gbaud data as the upper bound for nonlinear compensation capability at 160 Gbaud.

\begin{figure}[!htbp]
    \centering
    \includegraphics[width=0.48\textwidth]{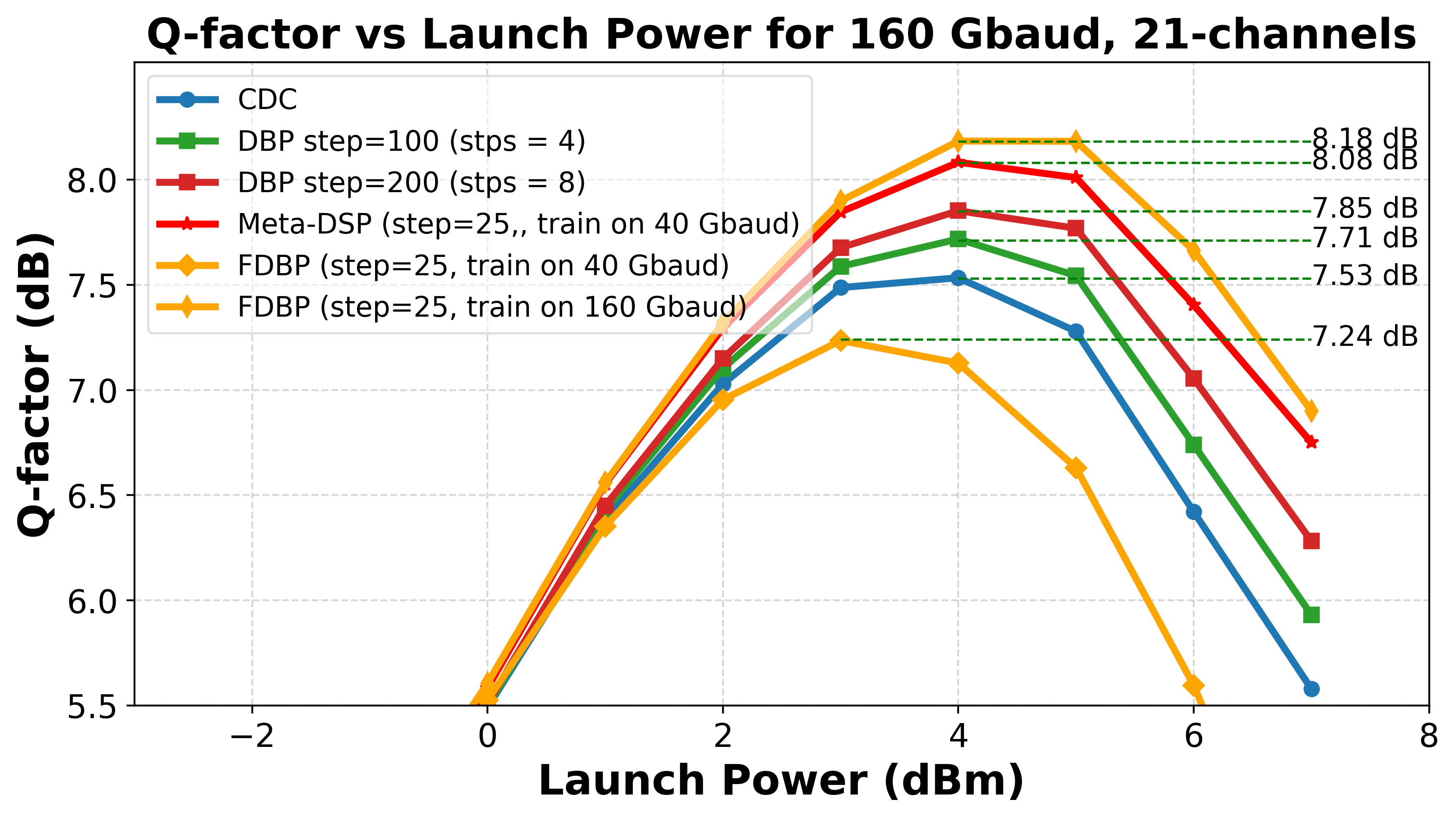}
    \caption{Q-factor vs. launch power for 160~Gbaud, 21 channels, and dual polarization. Meta-DSP, trained on 40~Gbaud data, achieves a Q-factor of 8.08~dB at 4~dBm, closely matching FDBP trained on 160~Gbaud data (8.18~dB) and outperforming CDC (7.53~dB) and DBP (7.71~dB and 7.85~dB). This demonstrates Meta-DSP's strong generalization and nonlinear compensation capabilities across symbol rates.}
    
    \label{Q-P-Nmodes2}
\end{figure}

As shown in Fig. \ref{Q-P-Nmodes2}, which presents the Q-factor performance at 160 Gbaud with 21 channels, the following results are observed. As baselines, CDC achieves the optimal Q factor of 7.53 dB at 4 dBm power, while DBP with steps=4 and steps=8 achieve optimal Q factors of 7.71 dB and 7.85 dB, respectively, at the same power level. Meta-DSP attains the optimal Q factor of 8.08 dB at 4 dBm power. In comparison, FDBP trained on 40 Gbaud data achieves an optimal Q factor of 7.24 dB at 3 dBm power, whereas FDBP trained on 160 Gbaud data reaches an optimal Q factor of 8.18 dB at 4 dBm power. It is evident that Meta-DSP, trained solely on 40 Gbaud data, exhibits remarkable Q-factor performance at 160 Gbaud with 21 channels, achieving a 0.55 dB gain over CDC and surpassing both DBP with steps=4 and steps=8. Furthermore, Meta-DSP's performance is nearly on par with that of FDBP trained on 160 Gbaud data, with only a marginal 0.1 dB difference. This suggests that Meta-DSP has effectively learned the majority of the nonlinear compensation capability required for 160 Gbaud from the low-speed 40 Gbaud data, highlighting its robust generalization ability. In contrast, FDBP trained on 40 Gbaud data underperforms compared to CDC at 160 Gbaud with 21 channels and dual polarization, demonstrating its inability to generalize to high-speed scenarios.

\begin{figure}[!htbp]
    \centering
    \begin{minipage}[b]{0.4\textwidth}
        \centering
        \includegraphics[width=0.9\textwidth]{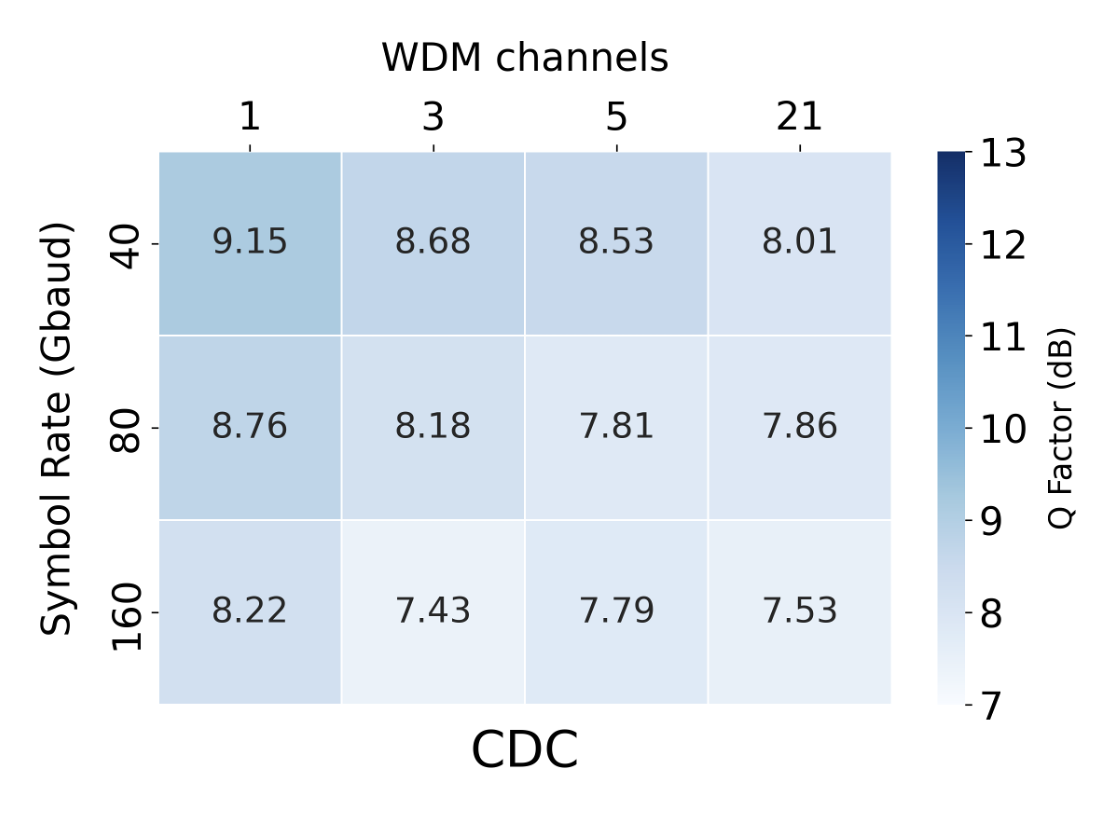}
        \caption*{(a) Optimal Q factor for CDC.}
    \end{minipage}
    \hspace{0.04\textwidth}
    \begin{minipage}[b]{0.4\textwidth}
        \centering
        \includegraphics[width=0.9\textwidth]{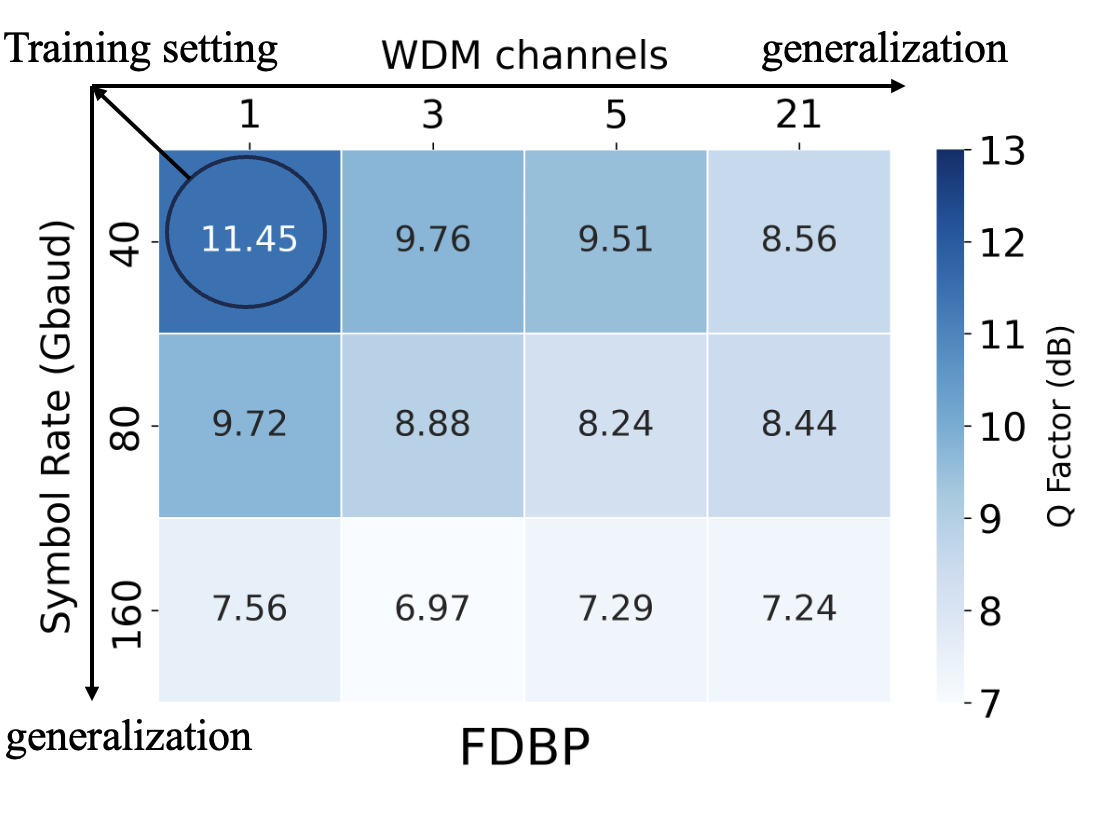}
        \caption*{(b) Optimal Q factor for FDBP (step=25) trained on 40G, 1 channel.}
    \end{minipage}
    
    \vspace{0.4cm}
    
    \begin{minipage}[b]{0.4\textwidth}
        \centering
        \includegraphics[width=0.9\textwidth]{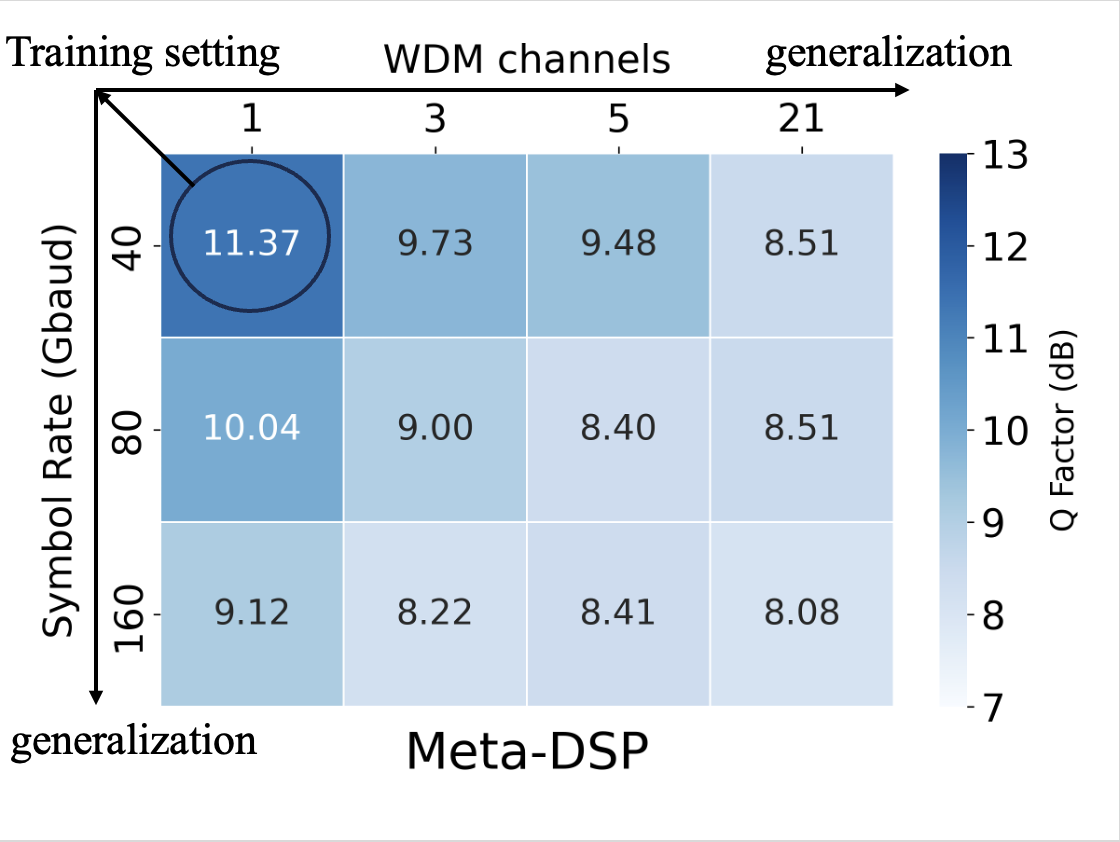}
        \caption*{(c) Optimal Q factor for Meta-DSP (step=25) trained on 40G, 1 channel.}
    \end{minipage}

    \caption{Optimal Q factor w.r.t power for compensation methods at different $R_s$, $N_{ch}$. We show Meta-DSP (step=25) and FDBP (step=25) trained on 40 Gbaud, 1 channel, dual polarization. Meta-DSP demonstrates strong knowledge transfer capabilities, significantly outperforming CDC across all settings. However, FDBP trained on a 40 Gbaud single channel does not possess the ability to transfer knowledge to high-speed environments.}
    \label{ex4}
\end{figure}

To provide a comprehensive overview of Meta-DSP's generalization capabilities, we present the optimal Q-factors with respect to the launch power $P_{ch}$ for various configurations of symbol rate $R_s$ and channel count $N_{ch}$ in Fig.~\ref{ex4}. As shown, Meta-DSP consistently outperforms FDBP in scenarios involving higher symbol rates. Notably, even in the most challenging case—dual-polarization at 160 Gbaud with 21 channels—Meta-DBP achieves a performance gain of 0.55 dB. While FDBP exhibits limited generalization when transferring from low to high symbol rates. This superior performance of Meta-DSP can be attributed to its ability to learn not just an N-filter specific to the training rate (e.g., 40 Gbaud), but a generalized functional form of the N-filter that applies across different symbol rates. When combined with a scale transformation described in Eq.~\eqref{N-filter}, this design enables the nonlinear compensation capability learned at 40 Gbaud to be effectively generalized to 80 Gbaud and 160 Gbaud.

\begin{table*}[htbp]
    \centering
    \begin{tabular}{|c|c|c|c|c|c|}
    \hline
    \textbf{Method} & \textbf{CDC} & \textbf{Meta-DBP step=5} & \textbf{FDBP step=5} & \textbf{Meta-DBP step=25} & \textbf{FDBP step=25} \\
    \hline
    Q Gain & 7.53 & 7.76 & 7.79 & 8.08 & 8.18 \\
    \hline
    Q Gain (over CDC) & 0.00\% & 3.06\% & 3.46\% & 7.30\% & 8.63\% \\
    \hline
    Time Consumption & - & < 0.1s (Hypernetwork) & >120s (SGD on new dataset) & < 0.1s (Hypernetwork) & >120s (SGD on new dataset) \\
    \hline
    \end{tabular}
    \caption{Comparison of Methods for Q Gain over CDC and Time Consumption. The test scenario is 160G with 21 channels. Meta-DSP uses a Hypernetwork trained on 40 Gbaud to generate parameters, which results in a much faster processing speed. In contrast, FDBP requires training with SGD and the collection of a new dataset when faced with a new scenario, leading to significantly higher time consumption. For both step=5 and step=25, Meta-DSP achieves performance close to that of FDBP trained on the new dataset.}

    \label{time_table}
\end{table*}

\begin{figure}[ht]
    \centering
    \includegraphics[width=0.52\textwidth]{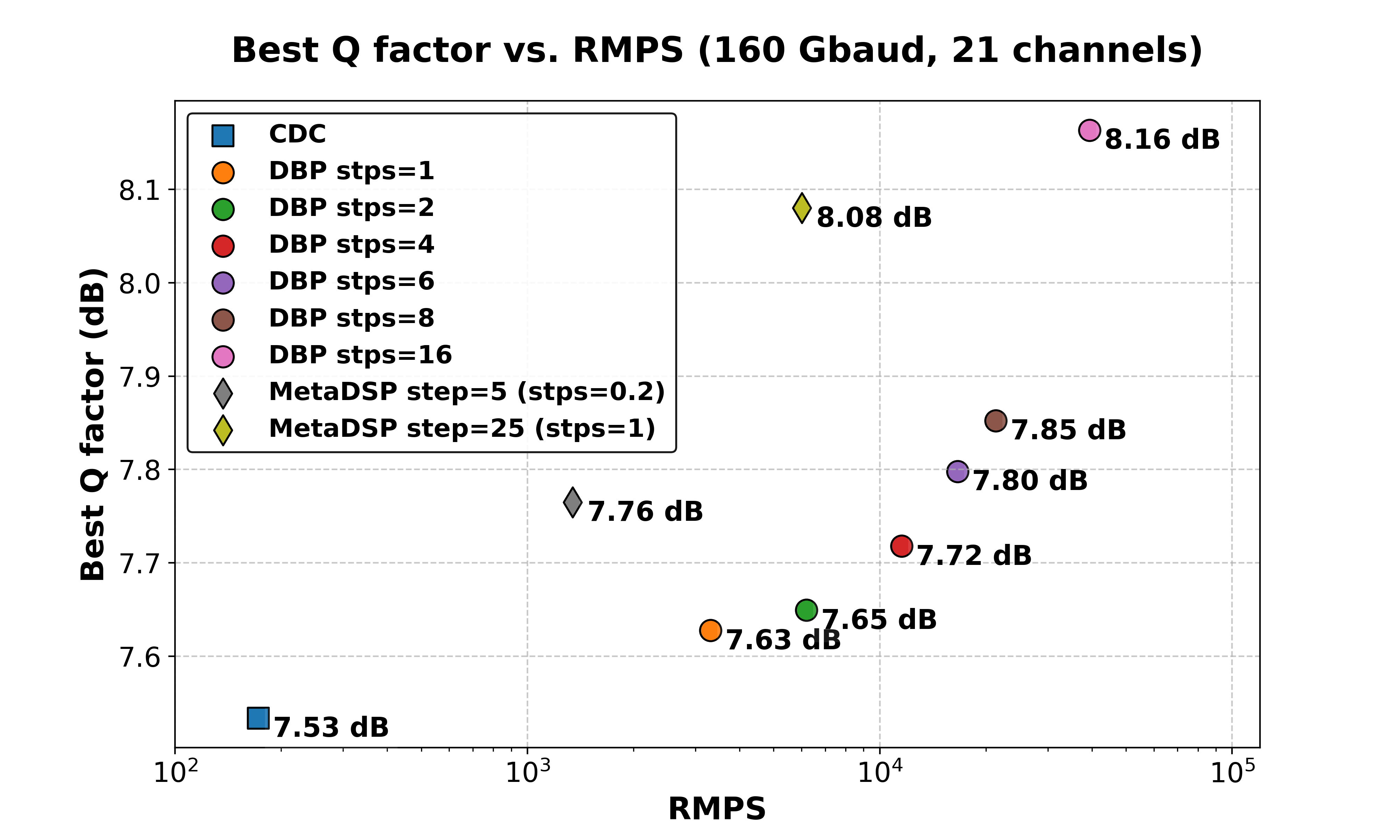}
    \caption{Comparison of computational cost and Q factor gains for various methods. We show results of 160 Gbaud, 21 channels, dual polarization. }
    \label{fig:cost_vs_gain}
\end{figure}

In comparison to traditional DBP methods, we evaluated the computational cost of various approaches relative to their Q factor gains, as illustrated in Figure \ref{fig:cost_vs_gain}. The results demonstrate that Meta-DSP reduces the computational cost to one-tenth of that of DBP while achieving equivalent Q factor performance. When compared to FDBP, which serves as the target network for Meta-DSP, Meta-DSP does not exceed FDBP in terms of ultimate performance. However, it exhibits a distinct advantage: despite being trained exclusively on low-speed data, Meta-DSP achieves performance comparable to FDBP in high-speed scenarios. This highlights its exceptional generalization capability, a feature fundamentally absent in FDBP. In our implementation, the hypernetwork is structured as a multilayer perceptron (MLP) with only 3 hidden layers. By leveraging highly optimized matrix multiplication on modern hardware, the MLP operates with high computational efficiency and minimal storage requirements. For fiber signal scenarios, the hypernetwork takes scenario-related metadata as input and rapidly generates model parameters, bypassing the need for complex training processes. As shown in Table \ref{time_table}, the inference speed of Meta-DSP's hypernetwork exceeds that of FDBP's time-consuming SGD training process by a factor of over 1000, making it a highly efficient and scalable solution for practical applications.

\newpage
\section{Conclusions}

In this study, we introduced Meta-DSP, a novel multi-modal nonlinear compensation algorithm for optical fiber receivers inspired by meta-learning. Meta-DSP demonstrates robust capabilities in processing optical fiber data under a myriad of conditions, including long-distance transmissions, varying symbol rates, multiple channels, and a range of power settings. Building on the concept of hyper-networks, we devised Meta-DBP to learn how to generate N-filter for nonlinear compensation, and developed XPM-ADF to learn how to compensate XPM.

Our numerical simulations evaluated Meta-DSP's performance on dual polarization 16-QAM signals. We tested transmission distances of 2000 km and explored symbol rates of 40 Gbaud, 80 Gbaud, and 160 Gbaud, along with WDM channel counts of 1, 3, 5 and 21. The transmission powers ranged from -8 dBm to 7 dBm. The results were encouraging: Meta-DSP trained at 40 Gbaud demonstrated strong generalization, successfully compensating for nonlinear effects even at higher rates and channel counts. The highest Q-factor improvements were 0.55 dB over CDC at 160 Gbaud with 21 channels.

Moreover, Meta-DSP achieved these gains while reducing computational complexity to one-tenth of that of DBP. These findings underscore the strong adaptability of Meta-DSP across various multi-modal scenarios and highlight the significant potential of data-driven paradigms in advancing nonlinear compensation algorithms for optical fibers.

Future research will focus on two key aspects. First, we will investigate the performance of Meta-DSP under more complex transmission models, such as the Manakov equation model with PMD. Second, we will explore the feasibility of Meta-DSP as a compensation algorithm under stronger optical power conditions. With the support of deep learning and data-driven modeling, our approach may pave the way to exploring the theoretical capacity limits of nonlinear channels.

\section*{Acknowledgement} Bin Dong is supported by the New Cornerstone Investigator Program and the National Natural Science Foundation of China
(No. 12288101).

%Bibliography
\newpage
\bibliographystyle{IEEEtran}
\bibliography{references} 
\end{document}